\documentclass[a4paper,12pt]{iopart}

\usepackage{graphicx,amssymb}

\def\Sc{Schr\"odinger }
\newcommand{\be}{\begin{equation}}
\newcommand{\ee}{\end{equation}}
\newcommand{\bq}{\begin{eqnarray}}
\newcommand{\eq}{\end{eqnarray}}

\begin{document}
\title{A Modified Borel Summation Technique} 

\author{David Leonard and Paul Mansfield}
\address{Centre for Particle Theory, Durham University, Durham, DH1 3LE, UK}
\eads{\mailto{david.leonard@durham.ac.uk},\mailto{p.r.w.mansfield@durham.ac.uk}}

\begin{abstract}
We compare and contrast three different perturbative expansions for the quartic anharmonic oscillator wavefunction and apply a modified Borel summation technique to determine the energy eigenvalues. In the first two expansions this provides the energy eigenvalues directly however in the third method we tune the wavefunctions to achieve the correct large $x$ behaviour as first illustrated in \cite{Leonard:2007dj}. This tuning technique allows us to determine the energy eigenvalues up to an arbitrary level of accuracy with remarkable efficiency. We give numerical evidence to explain this behaviour. We also refine the modified Borel summation technique to improve its accuracy. The main sources of error are investigated with reasonable error corrections calculated. 
\end{abstract}

%%%%%%%%%%%%%%%%%%%%%%%%%%%%%%%%%%%%%%%%%%%%%%%%%%%%%%%%%%%%%%%%%%%%
%%%%%%%%%%%%%%%%%%%%%%%%%%%%%%%%%%%%%%%%%%%%%%%%%%%%%%%%%%%%%%%%%%%%

\section{Introduction}
Quantum Field Theory (QFT) is our most successful mathematical framework for describing the fundamental laws of physics. Despite this we still have few tools to calculate physical quantities in many models. Perturbation theory is probably the most important technique available to calculate such quantities however perturbative expansions are often divergent within their region of applicability. In addition some physical properties are not correctly reflected in perturbative expansions. For example, the renormalisation group implies that energy eigenvalues in Yang-Mills theory cannot be solved for perturbatively. 

One technique that has been of interest in both the \Sc representation of quantum field theory and quantum mechanics is a modified type of Borel resummation \cite{Mansfield:1995gc}\cite{Mansfield:1995je}\cite{Mansfield:1997kj}\cite{man}.  The aim of this paper is to explore this particular method of resummation. It is important to test and develop the accuracy of our technique in a theory with known results so that we can confidently apply it to problems in which other techniques fail. Since energy eigenvalues are already known for the quantum mechanical anharmonic oscillator to a high degree of accuracy (e.g. \cite{Leonard:2007dj}\cite{kirsten} and references therein), we shall apply the resummation technique to this problem. That is we look for solutions to
\begin{eqnarray}
\label{aho}
-\frac{d^2\Psi}{dx^2}+\rho x^2\Psi+gx^4\Psi=E\Psi\\
\label{Psibc}
\lim_{|x|\to\infty}\Psi=0
\end{eqnarray}
where $x$ is defined along the real axis and we choose units in which $\hbar=2m=1$. We will generate three different perturbative expansions of the quartic anharmonic oscillator wavefunction as follows:
\begin{itemize}
\item An expansion in the coupling, $g$ (Bender Wu expansion)
\item An expansion in Plank's constant, $\hbar$ (the semi classical expansion)
\item An expansion in powers of $x$
\end{itemize}

In the first two approaches we take a perturbative expansion of the energy eigenvalue and directly resum to find an approximate result. In the third method we resum the large $x$ behaviour of the wavefunction. This provides a variational technique in which the energy is tuned to ensure the correct boundary condition \eref{Psibc} is observed. This technique is remarkably efficient and provides energy eigenvalues up to an arbitrary level of accuracy as first illustrated in \cite{Leonard:2007dj}. We shall apply the modified Borel resummation technique to provide further explanation as to why this method is so efficient.

%\label{directres}
The method of resummation we shall employ allows us to extract the small $s$ properties from an asymptotic expansion which is only valid for large $s$. Therefore consider an asymptotic expansion of a function $f(s)$ in inverse powers of $s$:
\begin{equation}
\label{asymf}
f(s)\approx a_0+\frac{c_1}{s}+\frac{c_2}{s^2}+\frac{c_3}{s^3}+\frac{c_4}{s^4}+\frac{c_5}{s^5}+\cdots.
\end{equation}
We analytically continue $f(s)$ into the complex $s$ plane, then Cauchy's theorem relates the large $s$ to small $s=s_0$ behaviour of $f(s)$ via the integral
\begin{equation}
\label{L(lambda)}
L(\lambda)=\int_C ds\, \frac{e^{\lambda\left(s-s_0\right)}}{s-s_0}f(s)
\end{equation}
where $C$ is a large circular contour centred on the origin. We only assume $f(s)$ to be analytic in the half plane $\Re(s-s_0)\ge 0$. Any singularity contributions to \eref{L(lambda)} in the half plane $\Re(s-s_0)<0$ are exponentially dampend by $\lambda$ so that $\lim_{\lambda\to\infty} L(\lambda)=f(s_0)$.

We approximate $L(\lambda)$ using a truncated version of the asymptotic expansion \eref{asymf} and expanding the $s-s_0$ denominator in powers of $s_0/s$ truncated to some order $P$. Thus we define
\begin{eqnarray}
\label{LN}
L_N(\lambda)&=\int_C ds\, e^{\lambda(s-s_0)}\sum_{p=0}^P\frac{s_0^p}{s^{p+1}}\sum_{n=0}^N\frac{c_n}{s^{n}}\\
&=\sum_{n=0}^N\sum_{p=0}^P c_n\frac{e^{-\lambda s_0}s_0^{p}\lambda^{n+p}}{\Gamma(n+p+1)} \approx L(\lambda).
\end{eqnarray}
where in completing the integral we used the identity $\int_C ds\,s^{-n}\exp(\lambda s)=2\pi i\lambda^{n-1}/\Gamma(n)$ for $n<0$. The introduction of the Gamma functions in this series improves the convergence of the original asymptotic expansion. We note however that the $L_N(\lambda)$ is only a good approximation to $L(\lambda)$ within a limited range of $\lambda$. For sufficiently large $\lambda$ the series will be dominated by the highest powers of $\lambda$ and exhibits a rapidly increasing or decreasing behaviour depending on the sign of the coefficient. We also note the similarity of this method to that of Borel summation. The Borel transform of an asymptotic series results in the introduction of an additional $1/n!$ factor in each $c_n$. The Borel procedure however requires us to analytically continue the Borel transformed series before inversion. This is accomplished via techniques such as Pad\'e approximants or conformal mapping. The advantage of our technique is that this analytic continuation is encoded in the contour integral.

We take $\lambda$ as large as practicably possible with the constraint that $L_N(\lambda)$ be a good approximation to $L(\lambda)$. This is best achieved by requiring maximal $\lambda=\lambda_M$ such that $L_N(\lambda)$ differs from $L_{N-1}(\lambda)$ by no more than a set amount. In this paper we will choose $\lambda_M$ so that they differ by no more than $10^{-3}$ percent. More terms (greater $N$) allows for a larger $\lambda$ and therefore better dampening of any singularity contributions.

One method for increasing the singularity dampening for a given $N$ is to introduce a new parameter, $\alpha$ by replacing $f(s)$ with $f(s^\alpha)$ and $s_0$ with $s_0^{1/\alpha}$ in $L(\lambda)$. For $\alpha>1$ the size of the last term in $L_N$ relative to its penultimate term $L_{N-1}$ is reduced due to the gamma function in \eref{LN}. This allows us to take a larger value of $\lambda$ whilst $L_N(\lambda)$ remains a good approximation to $L(\lambda)$. Increasing $\alpha$ however causes singularities of $f$ to be rotated about the origin. For $\alpha$ too large the singularities enter the half plane $\Re(s-s_0)\ge 0$ at which point they are no longer exponentially suppressed. $L(\lambda)$ will then exhibit oscillations resulting from these singularity contributions. We therefore take $\alpha$ to be as large as possible but still ensuring $L_N(\lambda)$ is monotonic as a function of $\lambda$ for $\lambda<\lambda_M$. The technique originally (i.e. without the introduction of $\alpha$) only worked for functions which are analytic in the half plane $\Re(s-s_0)\ge 0$. With the introduction of $\alpha$ we could also consider functions in which $f(s)$ has singularities with $\Re(s-s_0)\ge 0$. By reducing $\alpha<1$ we can rotate these singularities back into the half plane $\Re(s-s_0)<0$ where they become exponentially dampened.

In all three approaches we will solve the anharmonic oscillator ground state by writing $\Psi=e^{W}$ since the ground state of any quantum mechanical system has no nodes. In the case of the quartic anharmonic oscillator \eref{aho}, the potential and boundary condition are even in $x$. We therefore expand $W$ in the form $W=\sum_{n=1}^\infty a_n x^{2n}$. Substituting this expansion into the differential equation \eref{aho} and comparing coefficients of the $x^{2n}$ we get relations between the $a_n$ as follows
\begin{eqnarray}
\label{eq:a1-3}-2 a_1=E,&\quad\quad
-12 a_2-4a_1^2+\rho=0,&\quad\quad
-30 a_3-16a_1a_2+g=0
\end{eqnarray}
and for $n\ge 3$
\begin{equation}
\label{eq:rel}
a_{n+1}=-\left(\sum_{m=1}^n 4m(n-m+1)a_ma_{n-m+1}\right)/(2(n+1)(2n+1)).
\end{equation}
The equation \eref{eq:rel} allows us to find $a_{n+1}$ in terms of the $a_m$ with $m\le n$. We could therefore solve all of the $a_n$ in terms of $a_1$, $a_2$ and $a_3$. In turn these first three coefficients are determined by the physical parameters $E$, $\rho$ and $g$ via \eref{eq:a1-3}. However for a given $\rho$ and $g$ only specific values of $E$ allow the solution to satisfy the boundary condition \eref{Psibc}. In sections \ref{sec:gAHO}, \ref{sec:hbar} and \ref{tuning} we propose different methods for eliminating this final degree of freedom.

\section{Bender Wu Expansion}
\label{sec:gAHO}
Bender and Wu \cite{bender69} showed how to construct the ground state energy by summing all connected Feynman diagrams with no external legs. In general a Feynman diagram with $2n$ external legs has at least $n-1$ vertices. So we ensure that an $x^{2n}$ term is at least order $n-1$ in the coupling by making the expansion $a_n=\sum_{m=n-1}^\infty a_{n,m}g^m$. This is a similar approach to another method outlined by Bender and Wu in the same paper. We substitute this coupling expansion into the above relations between the $a_n$ and compare coefficients of $g^n$. The first coefficient is given by $a_{1,0}^2=1/4$ which requires a choice of sign for $a_{1,0}$. In keeping with the Bender Wu methods \cite{bender69} we choose a negative sign. In \cite{Leonard:2007dj} we showed that this sign choice  is required to ensure the correct boundary condition \eref{Psibc}.

Having made the choice for $a_{1,0}$, the remaining coefficients are uniquely determined. We first find each $a_{n,n-1}$ by looking at the $g^{n-1}$ coefficient in \eref{eq:a1-3} and \eref{eq:rel}. The $g^n$ coefficients give $a_{n,n}$ then the $g^{n+1}$ coefficients give the $a_{n,n+1}$ etc. At each stage we are substituting in the previous solutions. Eventually we can find each $a_n$ up to any order in $g$. Of course the coefficients will also depend on $\rho$ however for the purpose of this section we set $\rho=1$ without loss of generality since the eigenvalues for arbitrary $\rho$ may be reproduced via a form of Symanzik scaling \cite{Simon:1970mc}.

We now have a solution for the ground state wavefunction of \eref{aho} as the exponential of a power series in $x$ and $g$. The energy eigenvalue is computed from $a_1$ using \eref{eq:a1-3} as a power series in $g$. This gives the well known \cite{bender69}\cite{bender71}\cite{bender73} Rayleigh Schr\"odinger perturbation expansion for $E(g)$,
\begin{equation}
\label{eq:RS}
E=1+\frac{3}{4}g-\frac{21}{16}g^2+\frac{333}{64}g^3-\frac{30885}{1024}g^4+\frac{916731}{4096}g^5-\frac{65518401}{32768}g^6+\ldots.
\end{equation}

This expansion has a zero radius of convergence as can be seen from the asymptotic form of the $a_{1,n}$ coefficients at large $n$ as given by Bender Wu \cite{bender69}. It has been used to generate some energy eigenvalues via a Borel resummed Pad\'e approximants technique \cite{graffi}\cite{loeffel} although many other techniques have been used to find the energy eigenvalues more accurately and efficiently e.g. \cite{Leonard:2007dj}\cite{mb}\cite{halliday}\cite{Plo}\cite{Castro}\cite{Buenda}\cite{Sanchez}.

We will apply our resummation method to $E(g)$ in an attempt to get meaningful results for non zero values of $g$. To do this we analytically continue $g$ in the complex $s=1/g$ plane. We write $s_0=1/g$ and $c_n=-2a_{1,n}$ then apply the contour integral technique so that $L_N(\lambda)$ in \eref{LN} approximates the energy.
 
%\FIGURE
\begin{figure}
\begin{center}
\begin{tabular}{|c|c|c|c|c|c|}
\hline
$\mathbf{g}$ &$\mathbf{\alpha_M}$&$\mathbf{\lambda_M}$&$\mathbf{L_{30}}$&$\mathbf{E_\textit{best}}$& \bf{Error (\%)}\\
\hline
0.05783&1.6398&	2.9266&1.0397406&1.0397505&0.00095\\
0.13458&2.4250&7.6353&1.0846489&1.0846523&0.00031\\
0.23702&2.8159&10.4624&1.1359213&1.1359237&0.00021\\
0.37556&2.8160&9.9602&1.1952265&1.1952286&0.00017\\
0.56672&2.8160&9.6275&1.2649090&1.2649111&0.00017\\
0.83794&2.8159&9.3800&1.3483970&1.3483997&0.00020\\
1.23759&2.8160&9.1841&1.4509422&1.4509525&0.00071\\
1.85793&2.8160&9.0197&1.5810649&1.5811388&0.00467\\
2.89469&2.8160&8.8769&1.7536476&1.7541160&0.02671\\
4.83194&2.8160&8.7472&1.9974138&2.0000000&0.12948\\
9.19266&2.8160&8.6252&2.3766424&2.3904572&0.58127\\
23.50256&2.8160&8.5039&3.0767794&3.1622777&2.77882\\
206.09853&2.8160&8.3641&5.1098167&6.3245553&23.77265\\
\hline
\end{tabular}
\caption{Results for resummation of $E$ in the coupling.}
\label{fig:g}
\end{center}
\end{figure}
The results generated via our resummation method are listed in figure \ref{fig:g} and compared to the results as generated by the method \cite{Leonard:2007dj}. We label the eigenvalues generated by \cite{Leonard:2007dj} $E_\textit{best}$ since they are accurate to within the number of significant figures expressed. The seemingly strange choices for $g$ that we use becomes more natural in the semi classical expansion as outlined in the next section. We use the same values of $g$ in both sections to allow comparison.

The results for small $g$ are quite impressive with errors in the region of $10^{-3}$ to $10^{-4}$ percent however for larger $g$ the results are less impressive with the error approximately $23.8\%$ for the largest value of $g$. 

\begin{figure}
\begin{center}
$\begin{array}{ccc}
\includegraphics[scale=0.35]{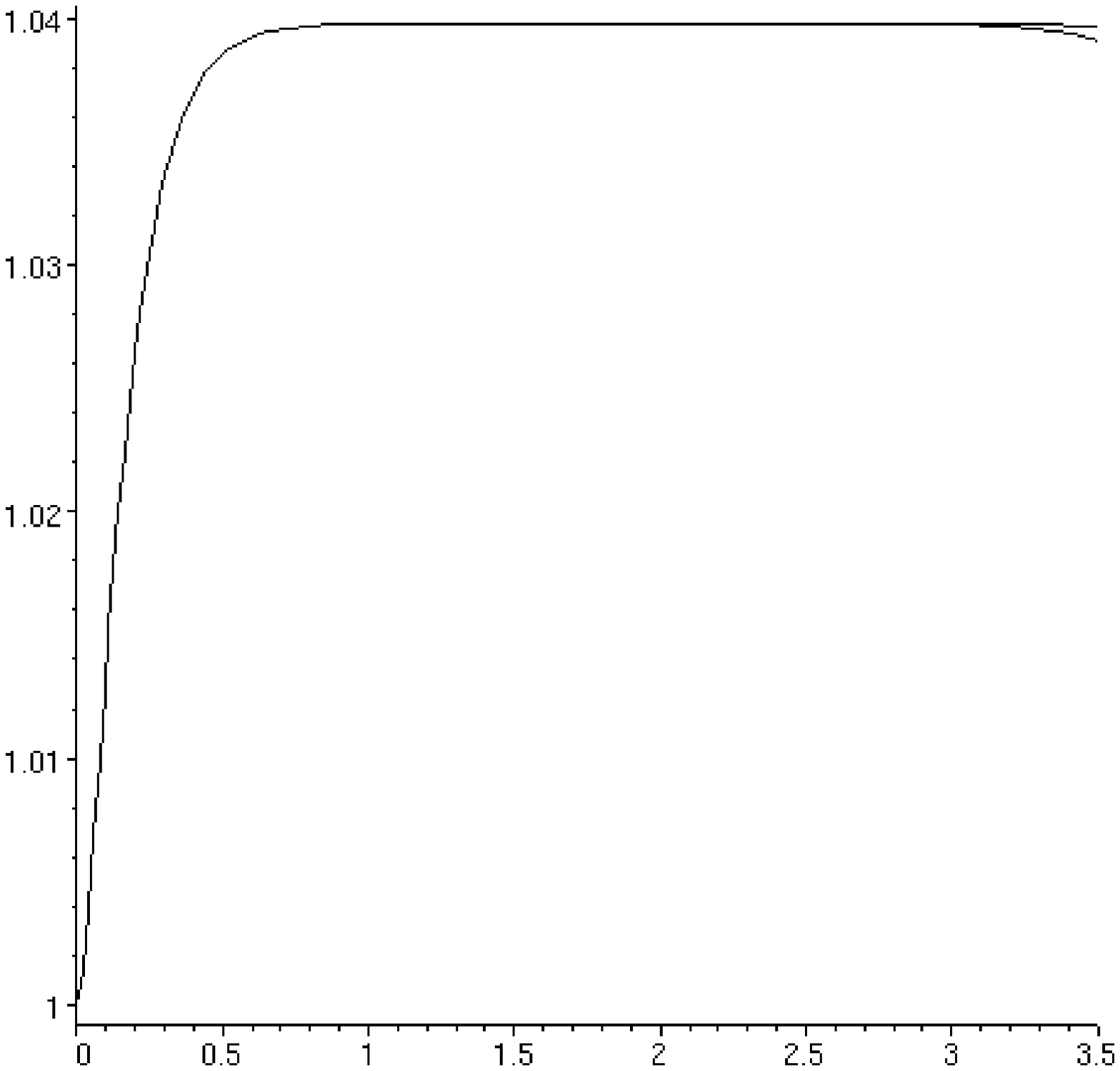}&\hspace{30pt}&
\includegraphics[scale=0.35]{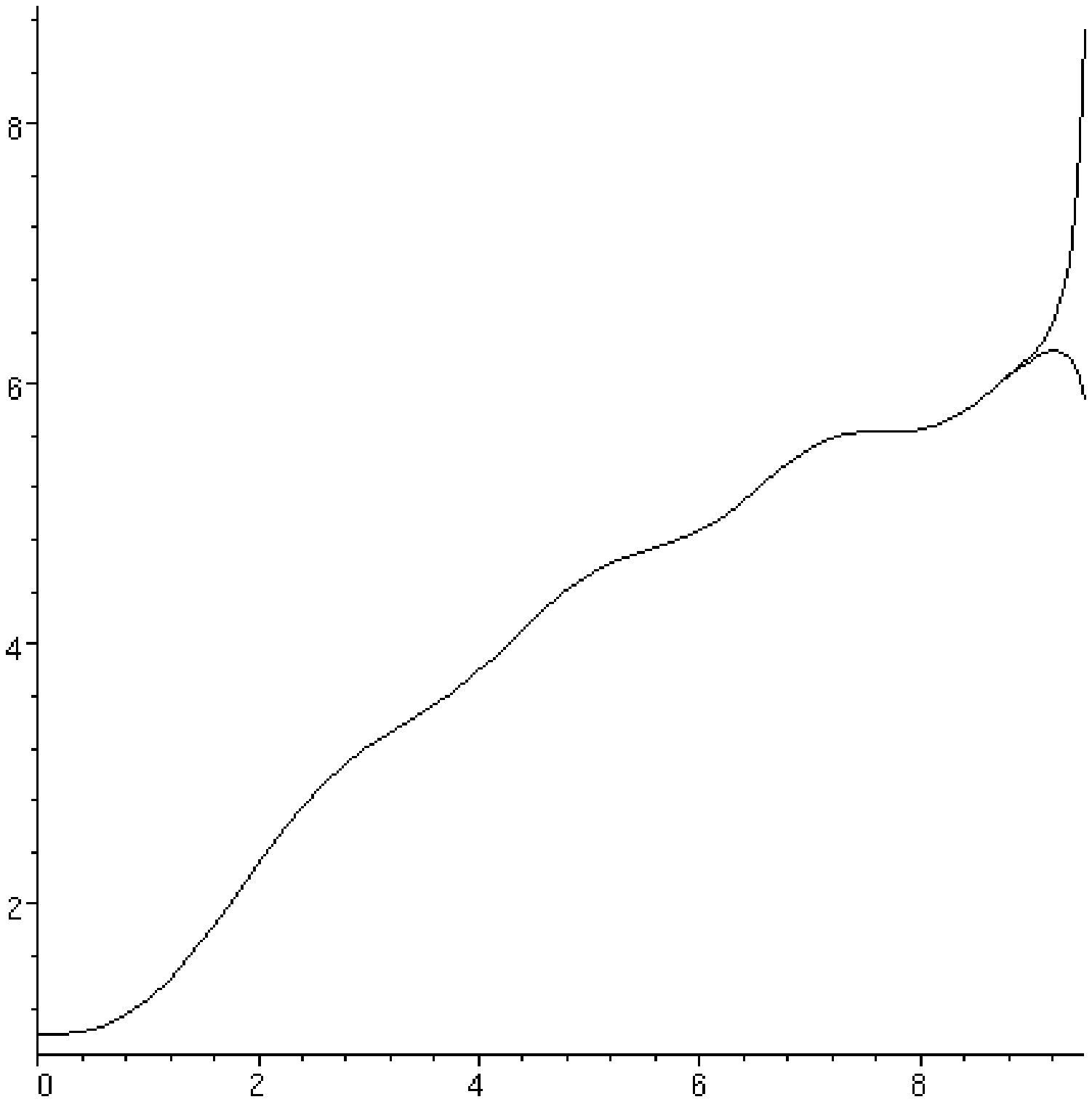}\\
$a) $g=0.05783&&$b) $g=206.09853
\end{array}$
\end{center}
\caption{Plot of $L_{30}(\lambda)$ and $L_{29}(\lambda)$ with $\alpha=\alpha_M$}
\label{fig:gl}
\end{figure}

Figure \ref{fig:gl}a (a small $g$ example) shows the expected behaviour of $L_{29}(\lambda)$ and $L_{30}(\lambda)$ with $P=50$. For sufficiently large $\lambda$ the two curves become a good approximation to $L(\lambda)$ and we see a flattening of the curve. For $\lambda$ sufficiently large we notice an appreciable divergence of the two curves. Figure \ref{fig:gl}b where $g$ is relatively large has somewhat different behaviour. Here we notice that the curves have not started to flatten before they appreciably diverge. For these larger $g$ we need more terms (greater $N$) in the expansion so that we may consider larger $\lambda$  where the curve starts to flatten. Additionally we note that this curve exhibits oscillatory behaviour although remains monotonic. When we introduced the requirement for the curve to be monotonic we assumed that a singularity contribution would consist of an exponentially weighted sinusoidal correction to a flat curve. In this case we do not have sufficient terms to consider $\lambda$ in the region where it becomes flat but instead are considering a region of the curve where it is still appreciably increasing. If we considered more terms with this value of $\alpha$ we may well find our monotonic condition is violated. Given the behaviour observed in figure \ref{fig:gl}b it is not surprising that we have such a large error.
%\FIGURE

In the next subsection we show that\footnote{This can be seen by Symanzik scaling of the Hamiltonian \cite{Simon:1970mc}. The infinite coupling limit has been calculated in \cite{parisi}.} $E\sim g^{1/3}$. Therefore as $g\to\infty$ or equivalently $s_0\to 0$ we notice that $E$ has a singularity. It is this singularity that is causing difficult in the resummation process since we need larger $\lambda$ for large $g$ to ensure it is damped sufficiently. This in turn requires a larger $N$. The problems no doubt could be solved if we took a sufficient number of terms in the expansion of $E$ however this would be at the expense of greater computing resources. In the next subsection we outline a more efficient method in which the singularity contribution is absent. We will therefore be able to calculate the infinite coupling limit. This is something that cannot be done via resummation of the Bender Wu expansion.

\section{Semi Classical Expansion}
\label{sec:hbar}
In this section we resum a semi classical expansion in $\hbar$ for the $a_n$ coefficients in $W$. We write $\Psi=e^{W/\hbar}$ and consider the modified differential equation
\begin{equation}
\label{eq:ahot}
-\hbar^2\frac{d^2\Psi}{dx^2}+\left(b_0+b_1x^2+b_2x^4\right)\Psi=0.
\end{equation}
The original problem \eref{aho} in which $\hbar=m=1$ is then recovered via a rescaling $x\to cx$ provided
\begin{eqnarray}
\label{eq:Eresc}
E=-b_0\frac{c^2}{\hbar^2},&\quad\quad\quad\quad\quad
\label{eq:mresc}
1=b_1\frac{c^4}{\hbar^2},\,&\quad\quad\quad\quad\quad
\label{eq:gresc}
g=b_2\frac{c^6}{\hbar^2}.
\end{eqnarray}
We substitute $\Psi$ into the $\hbar$ dependent differential equation to generate relations between the $a_n$. The equivalent expressions to \eref{eq:a1-3} are
\begin{eqnarray}
\label{eq:b}
b_0=2a_1\hbar,&\quad\quad\quad
b_1=4a_1^2+12a_2\hbar,&\quad\quad\quad
b_2=16a_1a_2+30a_3\hbar
\end{eqnarray}
and the new version of \eref{eq:rel} is
\begin{equation}
\label{eq:Wn}
2\hbar(n+1)(2n+1)a_{n+1}+\sum_{m=1}^n 4m(n-m+1)a_ma_{n-m+1}=0.
\end{equation}
The advantage of performing this rescaling is that we now have some freedom to choose $a_1$ and $a_2$. We will restrict the choice however by requiring $\hbar\ge 0$ with both $c$ and $\hbar$ real. This will allow us to choose $a_1$ and $a_2$ up to a sign. We will choose $a_1=-1/2$ and $a_2=-1/8$. We showed that this was the appropriate sign choice in \cite{Leonard:2007dj} to ensure that the boundary condition \eref{Psibc} is satisfied. With these choices we can reduce \eref{eq:Eresc} and \eref{eq:b} to
\begin{eqnarray}
\label{eq:E0}E=\frac{1}{\sqrt{1-\frac{3}{2}\hbar}},&\quad\quad\quad\quad\quad\quad
\label{eq:g}g=\hbar b_2 E^3.
\end{eqnarray}

We now assume an expansion $a_n=\sum_{m=0}^\infty a_{n,m} \hbar^m$ for each $n\ge 3$. This is substituted into \eref{eq:Wn} and coefficients of $\hbar^m$ compared for each $n$. We first compare coefficients of $h^0$ to get the $a_{n,0}$ then the coefficients of $\hbar^1$ give the $a_{n,1}$ etc. At each stage previous results are substituted into the new equation. By continuing this process sufficiently many times we can find each $a_n$ to any order required. A simple program can therefore be created to calculate $b_2$ (by using the $\hbar$ expansion of $a_3$) as an expansion in $\hbar$. The first few orders are
\begin{equation}
\label{eq:b2}
b_2=1+\frac{5}{8}\hbar-\frac{35}{32}\hbar^2+\frac{2555}{512}\hbar^3-\frac{69545}{2048}\hbar^4+\frac{4849705}{16384}\hbar^5-\frac{202337485}{65536}\hbar^6+\cdots
\end{equation}

\begin{figure}
\begin{center}\includegraphics[scale=0.5]{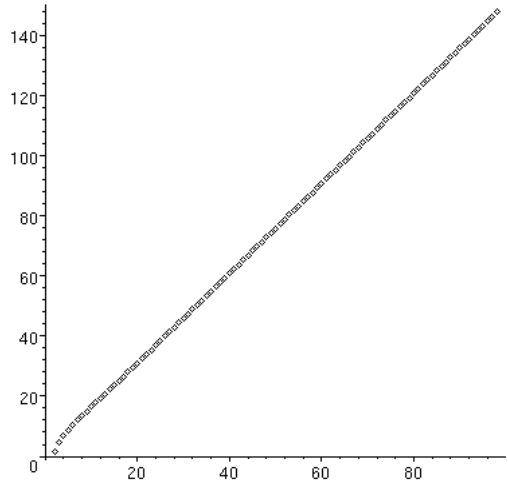}\end{center}
\caption{Ratio of coefficients of $\hbar^n$ in $b_2$}
\label{fig:hcoeff}
\end{figure}
The expansion of $b_2$ in $\hbar$ is an alternating sign series so we plot the ratio of successive coefficients of $\hbar^n$ in $b_2$ in such a way that we are dividing by the preceding term and removing the minus sign. We illustrate this in figure \ref{fig:hcoeff} for the first 100 coefficients and note the approximate linear behaviour for large orders. This suggests the asymptotic behaviour of the coefficients in $b_2$ have the form $(-1)^{n+1}k_1^n\Gamma(n+k_2)$ where $k_1$ and $k_2$ are real constants. Such asymptotic expansions have a zero radius of convergence as demonstrated via the alternating sign series test.

\begin{figure}
\begin{center}
\begin{tabular}{|c|c|c|c|c|c|c|}
\hline
$\mathbf{\hbar}$&$\mathbf{\alpha_M}$&$\mathbf{\lambda_M}$&$\mathbf{b_2}$&$\mathbf{g_{\textit{est}}}$&$\mathbf{g_\textit{best}}$&\bf{Err ($\mathbf{10^{-4}\%}$)}\\
\hline
0.05&1.5190&3.5078&1.0289799&0.0578315&0.0578320&8.85\\
0.10&2.2473&8.4878&1.0546589&0.1345810&0.1345815&3.68\\
0.15&2.5696&11.0496&1.0780575&0.2370176&0.2370183&2.69\\
0.20&2.7754&12.7628&1.0997450&0.3755562&0.3755570&2.20\\
0.25&2.8514&13.2331&1.1200769&0.5667191&0.5667201&1.83\\
0.30&2.8555&13.0189&1.1392958&0.8379415&0.8379430&1.71\\
0.35&2.8557&12.8266&1.1575774&1.2375925&1.2375945&1.66\\
0.40&2.8557&12.6736&1.1750540&1.8579236&1.8579267&1.65\\
0.45&2.8557&12.5499&1.1918288&2.8946853&2.8946902&1.70\\
0.50&2.8557&12.4457&1.2079838&4.8319353&4.8319442&1.83\\
0.55&2.8557&12.3584&1.2235859&9.1926365&9.1926555&2.07\\
0.60&2.8557&12.2820&1.2386905&23.5024991&23.5025564&2.44\\
0.65&2.8557&12.2158&1.2533439&206.0979166&206.0985278&2.97\\
\hline
\end{tabular}
\end{center}
\caption{Results of resummation in the semi classical expansion}
\label{tab:h}
\end{figure}
We shall therefore resum \eref{eq:b2} for a particular $s_0\equiv 1/\hbar$ from which $E$ and $g$ can be calculated. The $c_n$ in $L_N(\lambda)$ now correspond to the coefficients of $\hbar^n$ in the $b_2$ expansion. We note that $\hbar=0$ corresponds to $g=0$ and $E=1$ whilst $\hbar\to 2/3$ corresponds to $g\to\infty$. We actually find that $\hbar\in [0,2/3)$ corresponds to $g\in[0,\infty)$ which can be confirmed by the results of \cite{Leonard:2007dj}. We therefore calculate some couplings and energy eigenvalues within this range. The results are given in figure \ref{tab:h} again compared to results, $g_{best}$ (accurate to the stated number of significant figures) generated from \cite{Leonard:2007dj}. The energy eigenvalues produced are exact however it is the coupling that we are trying to approximate via the resummation process. We note that errors in the coupling are in the order of $10^{-4}\%$. Most importantly however the error remains within this order of magnitude for the full spectrum of $\hbar$ in contrast to resummation in the coupling. We can attribute this success to the fact that $\hbar\in[0,2/3)$ as opposed to $g\in[0,\infty)$ and the resummation process being most effective for small $\hbar$ or $g$. Also we encountered difficulties in resumming the coupling expansion for large $g$ due to the singularity at the origin in the $s$ plane. This problem has been removed in the semi classical expansion.

\begin{figure}
\begin{center}
$\begin{array}{ccc}
\includegraphics[scale=0.35]{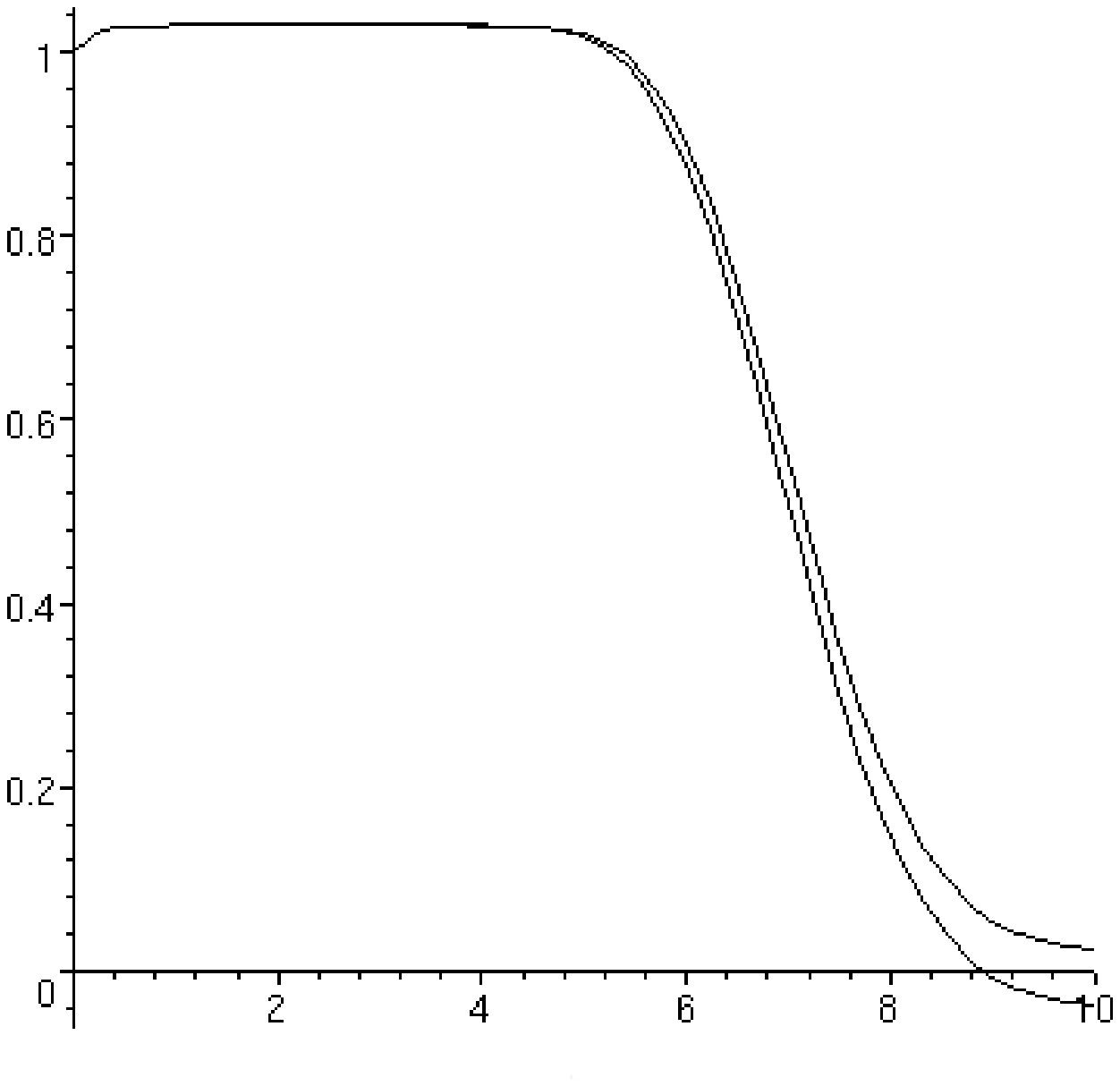}&\hspace{30pt}&
\includegraphics[scale=0.35]{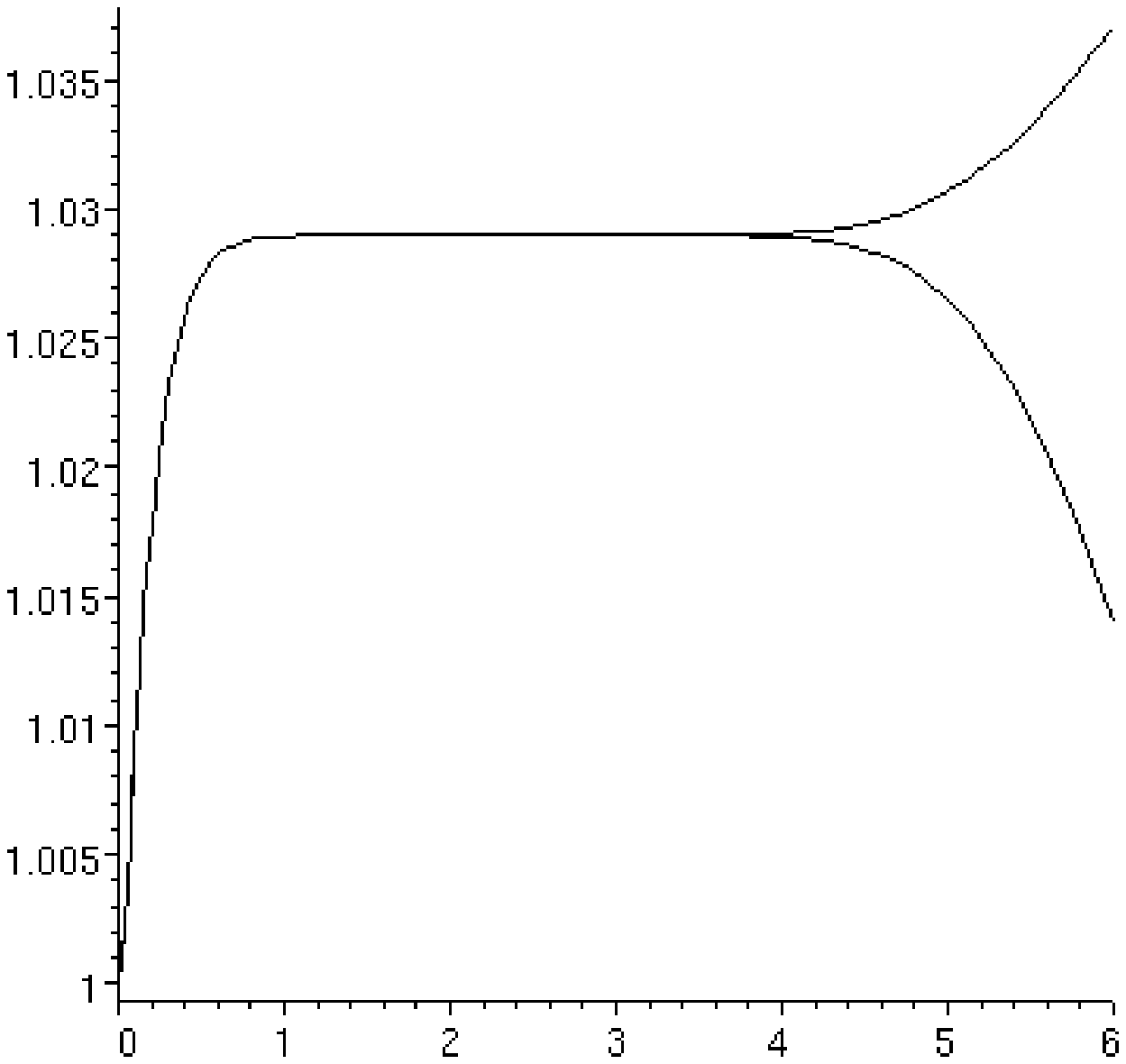}\\
$a) $P=50&&$b) $P=100
\end{array}$
\end{center}
\caption{Plot of $L_{30}(\lambda)$ and $L_{29}(\lambda)$ with $\hbar=0.05$}
\label{fig:hP}
\end{figure}

For higher $\hbar$ we expect the error to be greater since the contribution from singularities increases. It is interesting to note however that whilst this is true for the larger values of $\hbar$, the highest error is when $\hbar=0.05$. We attribute this to an insufficiently large value of $P$. We plot $L_{30}(\lambda)$ and $L_{29}(\lambda)$ in figure \ref{fig:hP}a with $P=50$ and note the decaying behaviour of the curves. With $P=100$ say we recover the expected behaviour of a flattening curve followed by the divergence of the two curves. One curve increases whilst one decreases from the point of divergence as illustrated in figure \ref{fig:hP}b. This is because the expansion of $(s-s_0^{1/\alpha})^{-1}$ in powers of $s_0^{1/\alpha}/s$ is only valid for large $s$. Whilst this is a valid assumption given the contour of integration, the series does require more terms to achieve a suitable level of approximation when $s_0$ becomes larger or equivalently $\hbar$ becomes smaller. The error as a result of truncation in this expansion is systematic hence the decaying nature of the curve for larger values of $\lambda$. This effect becomes more pronounced for larger $\alpha$. It is possible that $\alpha$ becomes sufficiently large to cause this decaying behaviour before singularity contributions becomes significant. However in this case the curve will still fail the monotonic condition. We resultingly take both a smaller $\alpha$ and a smaller $\lambda$ and therefore get greater singularity contributions than if we had taken a larger $P$. Despite the improvement expected with larger $P$ we are still able to extract good approximations with $P=50$.

The case $\hbar\to 2/3$ is particularly interesting because this corresponds to the infinite coupling limit. With $E_\infty=(3/b_2(2/3)/2)^{1/3}$ this limit corresponds to $E\to E_\infty g^{1/3}$. Using our resummation method we find $E_\infty=1.0603632150$ compared to the value $E_\infty=1.06036209$ which is exact to the stated number of significant figures. That is an error of approximately $1.06\times 10^{-4}\%$. Parisi \cite{parisi} was also able to calculate this limit as $E_\infty=1.06038$.

The method outlined so far for the ground state can be generalised to find the energy and wavefunctions of the excited states. We write the $q$th excited state $\Psi=P_q\Psi_0$ with energy $E=E_q+E_0$. Now consider
\begin{equation}
-h^2\frac{d^2\Psi}{dx^2}+(b_0+b_3+b_1x^2+b_2x^4)\Psi=0
\end{equation}
which using \eref{eq:ahot} reduces to
\begin{equation}
\label{eq:ahored}
-\hbar^2\frac{d^2P_q}{dx^2}-2W'\hbar\frac{dP_q}{dx}+b_3P_q=0.
\end{equation}
We scale $x\to cx$ to recover the AHO oscillator \eref{aho} provided that in addition to \eref{eq:E0} we also have
\begin{equation}
\label{eq:Eq}
E_q=-\frac{b_3c^2}{\hbar^2}=-\frac{b_3}{\hbar}E_0=-\frac{b_3}{\hbar\sqrt{1-\frac{3}{2}\hbar}}.
\end{equation}

We now need to find $b_3$ and $U_q$ using \eref{eq:ahored} which is easily done using a similar approach to that described for the ground state. That is expand
\begin{eqnarray}
U&=&\sum_{n=0}^\infty\sum_{m=0}^\infty c_{n,m}\hbar^mx^n,\\
b_3&=&\sum_{m=0}^\infty b_{3,m}\hbar^m,
\end{eqnarray}
and substitute into \eref{eq:ahored}. By comparing coefficients of $\hbar^k$ and $x^n$ we get
\begin{eqnarray}
\label{eq:exrel}
\eqalign{\fl-\sum_{i=0}^kb_{3,i}c_{n,k-i}+(n+2)(n+1)c_{n+2,k-2}\\
\eqalign+4\sum_{m=1}^{(n+1)/2}\sum_{i=0}^{k-1}m(n+2-2m)a_{m,i}c_{n+2-2m,k-i-1}=0}.
\end{eqnarray}
for $k>1$. The $k=0$ case reveals $b_{3,0}=0$ or $c_{0,n}=0$ for all $n$. Clearly the former is required. The $k=1$ case is given by
\begin{equation}
-b_{3,1}c_{n,0}+4\hbar\sum_{m=1}^{(n+1)/2}m(n+2-2m)a_{m,0}c_{n+2-2m,0}=0.
\end{equation}
We consider this equation for each $n$ starting with $n=0$ and find at each stage that $c_{n,0}=0$ until we reach a point where 
\begin{equation}
b_{3,1}=4na_{1,0}=-2n.
\end{equation}
Clearly $b_{3,1}=0$ is required for the ground state, $q=0$. Note also that a Taylor expansion of \eref{eq:Eq} would reveal $b_{3,1}=E_{q,0}$. This corresponds to the $q$th harmonic oscillator with energy eigenvalues $E=2q+1$ or $E_{q,0}=2q$. So we have $c_{n,0}=0$ for $n<q$ and $b_{3,1}=2q$.

We can fix one of the $x^n$ coefficients in $U_q$ by a choice of normalisation and therefore set $c_{q,0}=1$ with the remaining $c_{q,n}=0$. Now consider the $k=2$ case for each $n$ then $k=3$ etc. We see that at each stage \eref{eq:exrel} can be solved for $c_{n,k-1}$ provided $n\ne q$ or if $n=q$ we get $b_{3,k}$.

By solving a series of linear equations we have managed to determine $b_3$ and $U_q$ and hence found the excited wavefunctions and energies up to any order required. We can then apply our resummation method to evaluate the series expansion.

One advantage of this method is that we can calculate the energy eigenvalues for an infinite coupling. We have already seen that this is given when $\hbar=2/3$. By using \eref{eq:Eq} and \eref{eq:g} we can write
\begin{equation}
E_q=-\frac{b_3}{\hbar}\left(\frac{g}{\hbar b_2}\right)^{\frac{1}{3}}.
\end{equation}
So in the infinite coupling limit, $\hbar\to 2/3$ we have $E_{q}=E_{q,\infty}g^{1/3}$ where
\begin{equation}
E_{q,\infty}=-\frac{3}{2}b_3(2/3)E_{0,\infty}.
\end{equation}
Applying our resummation method we can calculate $b_3$ and hence $E_{q,\infty}$. A table listing the first ten $b_3$ is given in figure \eref{tab:Einf}. 
\begin{figure}
\begin{center}
\begin{tabular}{|c|c|c|c|c|c|}
\hline
$\mathbf{q}$&$\mathbf{\alpha_M}$&$\mathbf{\lambda_M}$&$\mathbf{b_3^\textit{est}}$&$\mathbf{b_3^\textit{best}}$&\bf{Error (\%)}\\
\hline
1&3.3554&17.9902&-1.722251&-1.722249&0.0001\\
2&3.2057&15.4576&-4.020691&-4.020850&0.0040\\
3&3.1254&13.8688&-6.654020&-6.654572&0.0083\\
4&3.0734&12.7075&-9.555955&-9.557404&0.0152\\
5&3.0341&11.7759&-12.681877&-12.686239&0.0344\\
6&3.0046&11.0209&-16.000874&-16.012209&0.0708\\
7&2.9831&10.4064&-19.489826&-19.514237&0.1251\\
8&2.9675&9.8993&-23.130233&-23.176133&0.1980\\
9&2.9560&9.4721&-26.906049&-26.984993&0.2926\\
10&2.9473&9.1068&-30.803510&-30.930247&0.4098\\
\hline
\end{tabular}
\end{center}
\caption{Infinite coupling values of $b_3$}
\label{tab:Einf}
\end{figure}

\section{Tuning the Boundary Condition}
\label{tuning}
In this section we use our resummation technique to try and explain why the method for generating energy eigenvalues in \cite{Leonard:2007dj} is so efficient. In this approach we examine the large $x$ behaviour of $W=\sum_{n=1}^\infty a_nx^n$ by applying the resummation technique already outlined. The differential equation implies large $x$ asymptotic behaviour of the form $W\sim\pm\sqrt{g}x^3/3$ and \eref{Psibc} then requires us to take the solutions that have a negative sign. We again employ Cauchy's theorem defining
\begin{equation}
\label{Tlambda}
T(\lambda)=\frac{1}{2\pi i}\frac{1}{\lambda^3}\int_C ds\,\frac{e^{\lambda s}}{s}W(s).
\end{equation}
so that when the boundary condition is satisfied we have $\lim_{\lambda\to\infty}T(\lambda)=-\sqrt{g}/18$ provided $W$ has no singularities in the right half $s\equiv 1/x$ plane. 

In the case of the quartic anharmonic oscillator, $W$ exhibits analytic behaviour in the whole complex $s$ plane however with higher order polynomial potentials this is not the case. A prescription in which we take $W(s)\to W(s^\alpha)$ and $1/\lambda^3\to1/\lambda^{3\alpha}$ in \eref{Tlambda} was introduced. By reducing $\alpha<1$ singularities are rotated into the half plane $\Re(s)<0$ in which case they become exponentially suppressed. Since we will restrict ourselves to the quartic oscillator in this paper we shall take $\alpha=1$.

A rescaling $x\to cx$ ($c\in\mathbb{R}$) in the differential equation allows us to fix the ratio $k\equiv a_1/a_2=\pm 4$. We make a choice for $a_2$ and sign of $k$ then solve the remaining $a_n$ in terms of $a_3$ via the recurrence relation \eref{eq:rel}. We vary $a_3$ until we observe the correct boundary condition in $T(\lambda)$. As with the previous direct resummation methods we use a truncated expansion of $W$ to approximate $T(\lambda)$ with $T_N(\lambda)$. 

When $T_N(\lambda)$ is plotted for $a_3$ too small to satisfy the boundary condition, we find a rapidly decreasing curve. If $a_3$ is too large then we find a rapidly increasing curve. We are therefore able to tune $a_3$ by adjusting upper and lower bounds to find an interval within which $a_3$ lies. We start with a modest value of $N$ and tune $a_3$ until $T_N(\lambda)$ produces a curve flattening for the larger values of $\lambda$. The range of $\lambda$  considered should be chosen to ensure $T_N$ is a good approximation to $T$. That is up to a value of $\lambda$ at which $T_N$ and $T_{N-1}$ appreciably diverge. To further increase the level of accuracy we need to consider whether the curve is rapidly increasing or decreasing for larger values of $\lambda$. We therefore need to increase $N$. So $N$ effectively determines the level of accuracy. 

With the sign choice of $k$ and value of $a_2$, $a_3$ is determined up to an arbitrary level of accuracy. The physical parameters are then found via 
\begin{eqnarray}
\label{Em}
E=\frac{-2ka_2}{(16ka_2^2+30a_3)^{\frac{1}{3}}}g^{\frac{1}{3}}\\
\rho=\frac{4k^2a_2^2+12a_2}{(16ka_2^2+30a_3)^{\frac{2}{3}}}g^{\frac{2}{3}}.
\end{eqnarray}

The procedure can be generalised to find the excited energy eigenvalues. As with the semi classical expansion we write the $q$th excited state $\Psi_q=P_q\Psi_0$ and energy $E=E_0+E_q$. Then $P_q$ in expanded in powers of $x$, $P_q=\sum_{n=0}^\infty c_nx^n$. The even powered coefficients are set to zero for $q$ odd and the odd powered coefficients to zero for $q$ even. 
Again we used a scaling of the differential equation $x\to cx$ and substitute in $\Psi_q$. We compare coefficients of $x^n$ to get a recurrence relation for the $c_n$. This also determines $E_q$ in terms of the $c_n$. We set the first non zero coefficient in the	 expansion ($c_0$ or $c_1$) to unity and use the recurrence relation to solve the $c_n$ in terms of $c_2$ or $c_3$. We now define $U$ via 
\begin{equation}
U(\lambda)=\frac{1}{2\pi i}\int_C ds\,\frac{e^{\lambda s}}{s}P_q(s^\alpha)
\end{equation}
and use the truncated expansion for $P_q$ to approximate $U(\lambda)$ with $U_N(\lambda)$. 

We vary either $c_2$ or $c_3$ as before and again observe the rapidly increasing or decreasing behaviour for solutions which do not correspond to eigenstates. For example in the odd case we vary $\tau=-c_3$. There will be multiple values of $\tau$ which correspond to different levels of excitation. We label these $\tau_n$ such that $\tau_{n+1}>\tau_n$. With $\tau<\tau_1$. $U(\lambda)$ corresponds to a rapidly increasing curve and with $\tau_1<\tau<\tau_2$ we get a rapidly decreasing curve. With $\tau_2<\tau<\tau_3$ $U(\lambda)$ switches back to a rapidly increasing curve. This switch between rapidly increasing and rapidly decreasing occurs whenever we increase $\tau$ past one of the $\tau_n$. The level of truncation determines the level of accuracy as before. 

Although the ground state was analytic in the whole complex $s$ plane for the quartic oscillator we found that the excited states were not. Therefore when we tune the curve, we get oscillations instead of a flattening behaviour. We could in principal use our technique of decreasing $\alpha<1$ to move the poles into the left half plane where they are suppressed by the exponential factor. This reproduces the flattening behaviour of the $T_N(\lambda)$ curve. We find however that it suffices to tune for the oscillations. 

One of the advantages of this approach compared to direct resummation of a perturbative expansion in $\hbar$ or $g$ is that we are able to produce solutions in which $\rho<0$. These solutions are dominated by instanton effects which are non perturbative in both $\hbar$ and $g$. Also this process allows us to determine an energy eigenvalue to an arbitrary level of accuracy. It is the switch between the rapidly increasing or decreasing behaviour of $T_N$ or $U_N$ that makes this method work so well. In \cite{Leonard:2007dj} we postulated a reason for this and in this section we use our resummation techniques to further investigate this behaviour. 

\subsection{The Ground State - Zeros, Poles and Cuts}
\label{zeros}
If we solve the differential equation \eref{aho} for its large positive $x$ behaviour without the boundary condition \eref{Psibc} we find two asymptotic solutions, $W=\pm\sqrt{g}x^3/3$. The general large $x$ solution to \eref{aho} is therefore of the form
\begin{equation}
\Psi_l=\exp\left({-\frac{\sqrt{g}}{3}x^3}\right)+A\exp\left({\frac{\sqrt{g}}{3}x^3}\right).
\end{equation}
The boundary condition \eref{Psibc} however requires us to take $A=0$. $\Psi_l$ in general has zeros at specific points in the complex $x$ plane depending on $A$. The only exception to this is when $A=0$ at which point we only asymptotically approach zero as $x\to\infty$. For $A<0$ these zeros lie along the real axis whereas for $A>0$ they lie off the real axis, somewhere in the complex $x$ plane. There will also be contours in the complex $x$ plane along which $\Psi_l$ is purely real and negative. The zeros and negative regions in $\Psi_l$ will be manifested as cuts and poles in $\log\Psi_l$. 

%We should note however that we construct $\Psi=\exp(W)$ and it is possible to construct an analytic $W$ which results in negative regions for $\Psi$ whereas $\log(\Psi)$ would have cuts. So whilst a zero in $\Psi$ necessarily corresponds to a pole or part of a cut in $W$ it is not necessarily true that a negative region in $\Psi$ corresponds to a cut in $W$. The $A=0$ case of $\Psi_l$ is an example of this in that $\Psi_l$ may have regions in which it is purely negative but the corresponding $W_l$ is an analytic function. 

It is clear that any pole or cut in the right half complex $x$ plane would result in an exponentially increasing or decreasing $L_N(\lambda)$. Oscillations will occur due to a pole lying off the real axis however we took modest $N$ to ensure we only see the beginning of the oscillation and hence the appearance of a rapidly increasing or decreasing curve. We hypothesised in \cite{Leonard:2007dj} that this is what is being observed in $L_N(\lambda)$. In this section we present numerical evidence to support this. We are essentially tuning the solution until $A=0$. For $A\ne0$ the resummation is conveniently spoilt in such a way that we are able to refine the solution. In this section we will employ Cauchy's theorem to determine the location of zeros in the solutions to the differential equation as constructed in the previous section. We will then be able to observe the dependence of these zeros on $a_3$. Whilst we have so far only presented an argument based on the large $x$ behaviour of $\Psi$ we will find that the location of zeros in $\Psi_l$ do indeed correspond approximately to the location of zeros in the full solution. 

We note that since the coefficients $a_n$ in $W$ (and a Taylor expanded $W_l$) are all real, any cut or pole in the upper half $s$ plane should necessarily be mirrored in the lower $s$ plane. Let us momentarily assume that in the right half $s$ plane $W(s)$ only contains poles located at $s_p$ and $s_p^*$. We then construct
\begin{eqnarray}
%\begin{split}
\eqalign{S(\lambda)&=\frac{1}{2\pi i}\int_C ds\,\frac{e^{\lambda (s-s_0)}}{s-s_0}W(s)\\
&\sim W(s_0)+c\cos(\lambda s_p^\Im+\nu)e^{\lambda(s_p^\Re-s_0)}}
%\end{split}
\end{eqnarray}
where $c$, $\nu$ are real numbers and $s_p$ is split into real and imaginary parts, $s_p^\Re+is_p^\Im$. If we had included a cut contribution instead of a pole then we would expect contributions from the whole cut with each portion of the cut weighted by the exponential factor. In $\Psi_l$ the zero lies on the right end of a region of negativity corresponding to a potential cut in $W_l$ and therefore is the dominant singularity contribution in our contour integral if $W=W_l$. So whilst $W(s)$ does not necessarily have just one pole in the positive right quadrant we will model the cuts by a pole representing a weighted average. Since the zero is the dominant contributor we will find that $S(\lambda)$ at least in the case of $W_l$ allows us to reasonably approximate the location of our zero. We will apply this model to our original $\Psi$ and predict zeros close to the zeros of $\Psi_l$. As usual we use the truncated expansion of $W$ to approximate $S(\lambda)$ by $S_N(\lambda)$.

With the above construction we are in a position to numerically determine the location of the conjugate poles. Initially we will assume that $s_p^\Im\ne 0$ and plot $S_N(\lambda)$ for a given $s_0$. We expect to see oscillations in the curve either growing ($s_p^\Re>s_0$), decaying ($s_p^\Im<s_0$) or fixed in amplitude. We tune $s_0$ until we achieve the latter at which point we have $s_p^\Re=s_0$. With this $s_0$ we can determine the frequency of the oscillations in $S_N(\lambda)$, this gives us $s_p^\Im$.  

If we do not observe the oscillatory behaviour in $S_N(\lambda)$ even with $N$, $P$ sufficiently large then it may be that $s_p^\Im=0$. If this is the case then we use $S_N(\lambda)$ to approximate $W(s_0)$ for $s_0>s_p$. As $s_0$ becomes closer to $s_p$ however the effect of our pole becomes more significant since the exponential dampening factor becomes reduced. We must therefore take larger $\lambda$ which in return requires larger $N$. Alternatively we can try increasing $\alpha>1$. This causes the final term in $S_N$ to be reduced in comparison to the penultimate term due to the $\Gamma$ function. This again allows a larger $\lambda$ and hence greater exponential dampening. Ultimately though we are unable to extract $W(s_0)$ from $S_N(\lambda)$ for $s_0\le s_p$ and indeed the process becomes more difficult as $s_0$ approaches $s_p$. We can however determine $s_p$ by approximating $\Psi$ to a linear behaviour in this region. We justify this by solving the differential equation in a region where $W'$ and $W''$ are more dominant than any potential term. Numerically this linear behaviour appears to work well at least within the region that we were realistically able to plot.

\begin{figure}
\begin{center}
\includegraphics[scale=0.5]{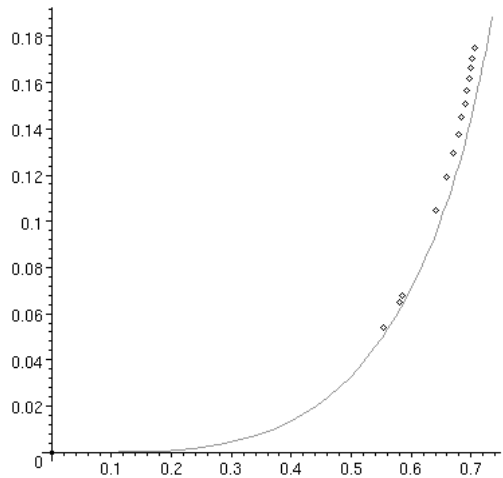}
\end{center}
\caption{Location of zeros}
\label{pzeros}
\end{figure}

For $a_3$ too small we find that the zeros lie along the real $s$ axis with zeros approaching $s=0$ (i.e. $x=\infty$) and indeed attaining this value as $a_3$ becomes appropriate for the correct boundary condition to be satisfied. For $a_3$ too large we find that the location of zeros lie off the real axis and into the complex plane. They appear to lie on some countour which approaches $s=0$ as $a_3$ approaches its correct value. We plot the location of some of these complex zeros of $\Psi$ in 
figure \ref{pzeros}\footnote{We have actually scaled the complex plane so that  $s^3\to \sqrt{16a_1a_2+30a_3}s^3/3$ for simplicity of calculation in figure \ref{pzeros}}. We have restricted this plot to one quadrent only however the plot would  
necessarily be mirrored into all four quadrants. The solid line represents the location of zeros of $\Psi_l$ as constructed at the beginning of this section. 

We note that the two sets of zeros are not in exact agreement although there does appear to be a correlation. We attribute the differences largely due to $\Psi_l$ being a small $s$ approximation and we therefore expect the approximation to improve as $|s|\to0$ which appears to be happening in the plot. Unfortunately it becomes increasing difficult to numerically calculate zeros of $\Psi$ as we get closer to the origin since the period of oscillation becomes very large. In order to measure this period we need to be able to plot at least the first few oscillations. This requires an increasingly large $\lambda$ which in turn increasingly requires a large number of terms.

We also question how accurately our numerical technique can accurately determine the location of zeros. As an example we calculate analytically the location of the zeros in $\Psi=\exp(x^3)+\exp(-x^3)$. Applying the resummation process to $\log(\Psi_t)$ expanded in positive powers of $x$ we are able to numerically calculate the location of the zero. We find analytically that a zero exists at $0.745+0.43i$ whereas numerically we find it exists at $0.738+0.429i$. Given the scale of the plot this error is relatively insignificant.

\subsection{Excited States - Large $x$ behaviour}
\label{excited}
We substitute $\Psi_q=P_q\Psi_0$ into the differential equation \eref{aho} to produce a new differential equation satisfied by $P_q$,
\begin{equation}
\label{Pdiff}
\frac{d^2P}{dx^2}+2\frac{dW}{dx}\frac{dP_q}{dx}+E_qP_q=0.
\end{equation}
Taking $W\sim -\sqrt{g}x^3/3$, this differential equation exhibits two types of large $x$ asymptotic solution
\begin{equation}
\label{largeP}
P\sim \exp{\left(-\frac{E_q}{2\sqrt{g}x}\right)}\qquad\mbox{or}\qquad P\sim \exp{\left(\frac{2\sqrt{g}x^3}{3}\right)}.
\end{equation}
We necessarily choose the first of these to ensure the boundary condition \eref{Psibc} is satisfied. When the boundary condition is satisfied $U_N(\lambda)$ produces a flat curve (having adjusted $\alpha$ appropriately) which would be expected if we have correctly chosen the large asymptotic behaviour. If the second type of asymptotic behaviour is chosen then we would expect to observe a correction to $T_N(\lambda)$ due to a singularity at $s=0$. We hypothesised in \cite{Leonard:2007dj} that when $c_2$ or $c_3$ does not correspond to an energy eigenstate then we are observing the second type of asymptotic large $x$ behaviour. 

\begin{figure}
\begin{center}
$\begin{array}{ccccc}
\includegraphics[scale=0.28]{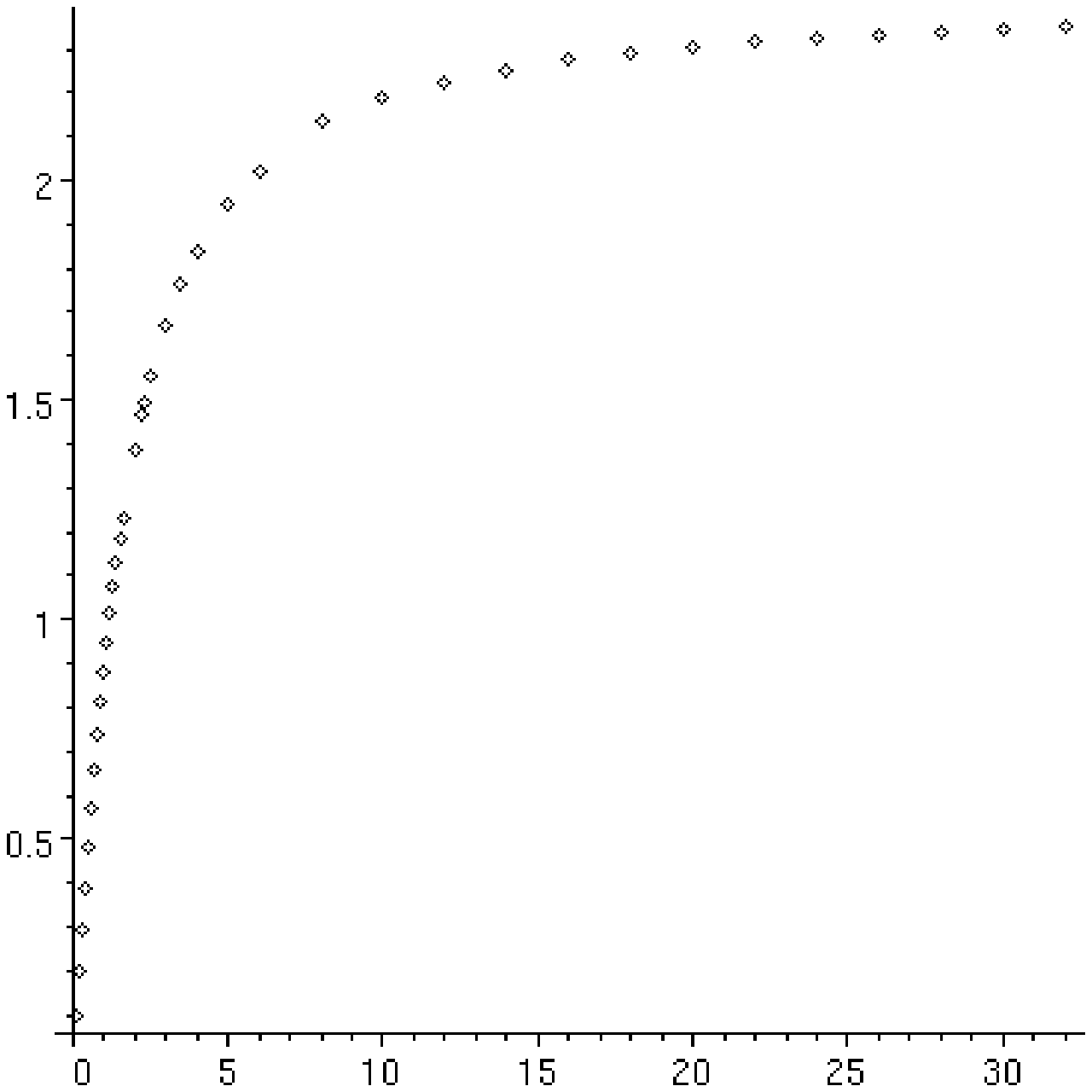}&\hspace{7pt}&\includegraphics[scale=0.28]{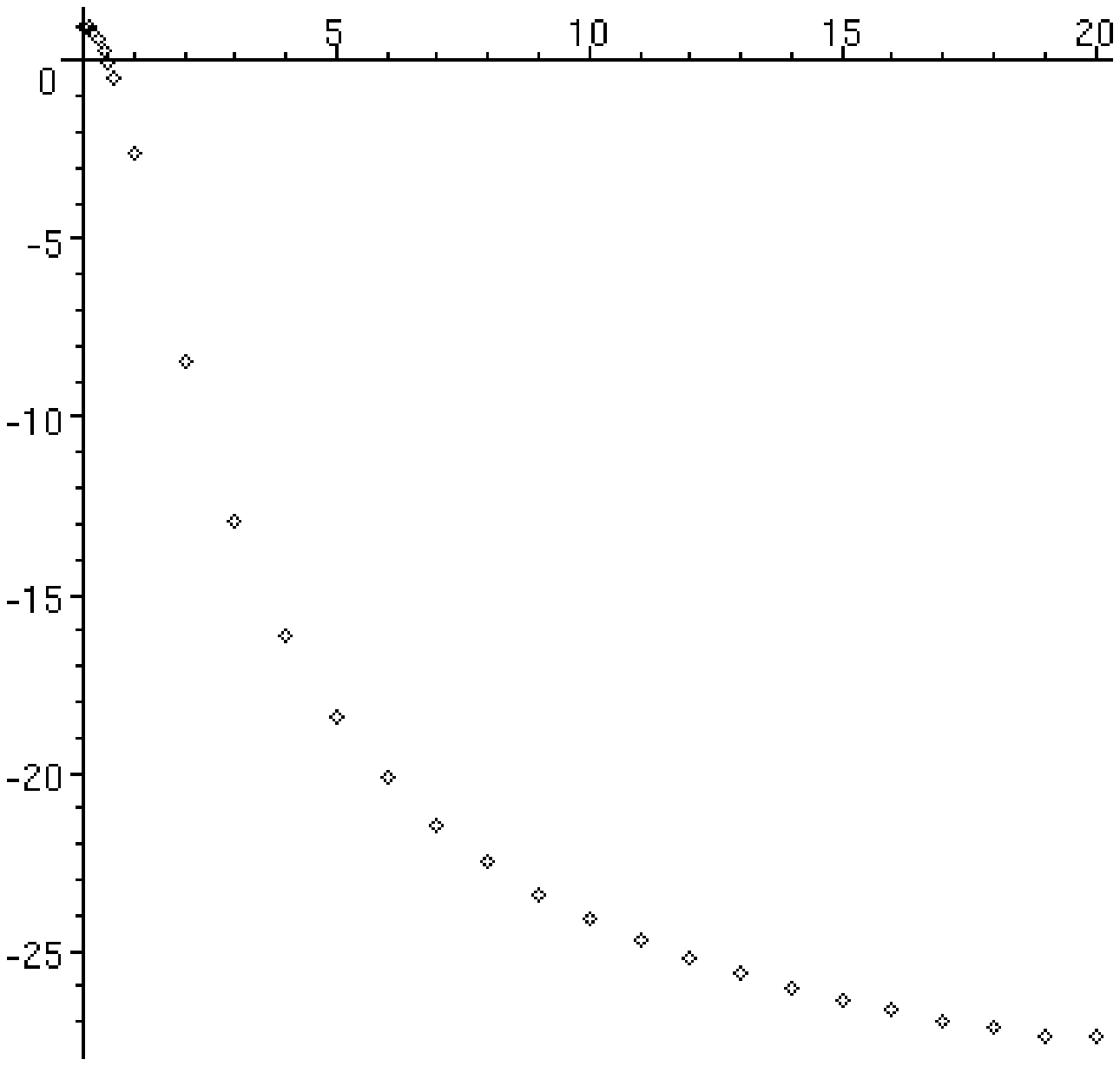}&\hspace{7pt}&\includegraphics[scale=0.28]{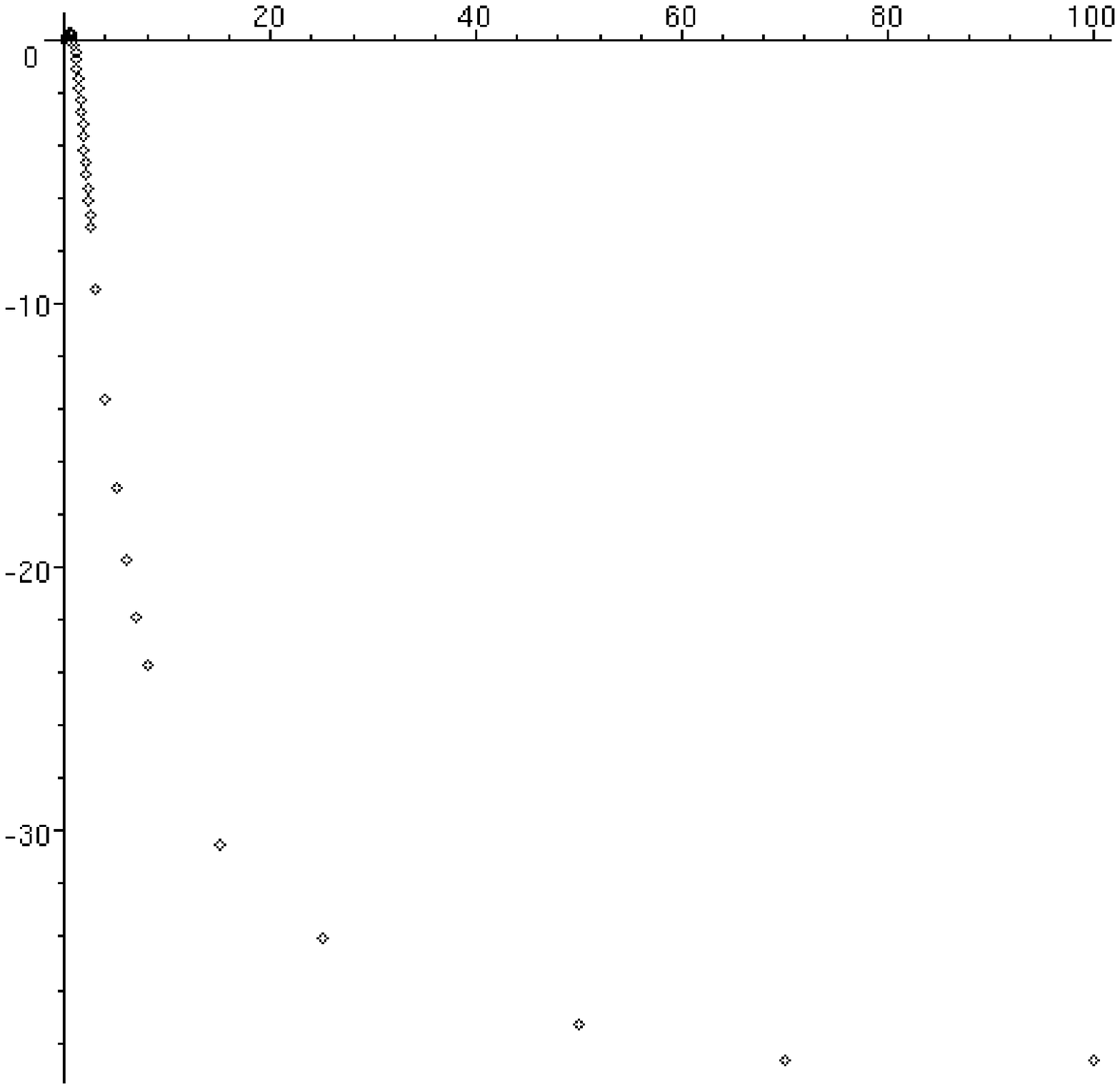} \\
$a) $P_1&&$b) $P_2&&$c) $P_3\\
\vspace{7pt}
&\hspace{7pt}		&\includegraphics[scale=0.28]{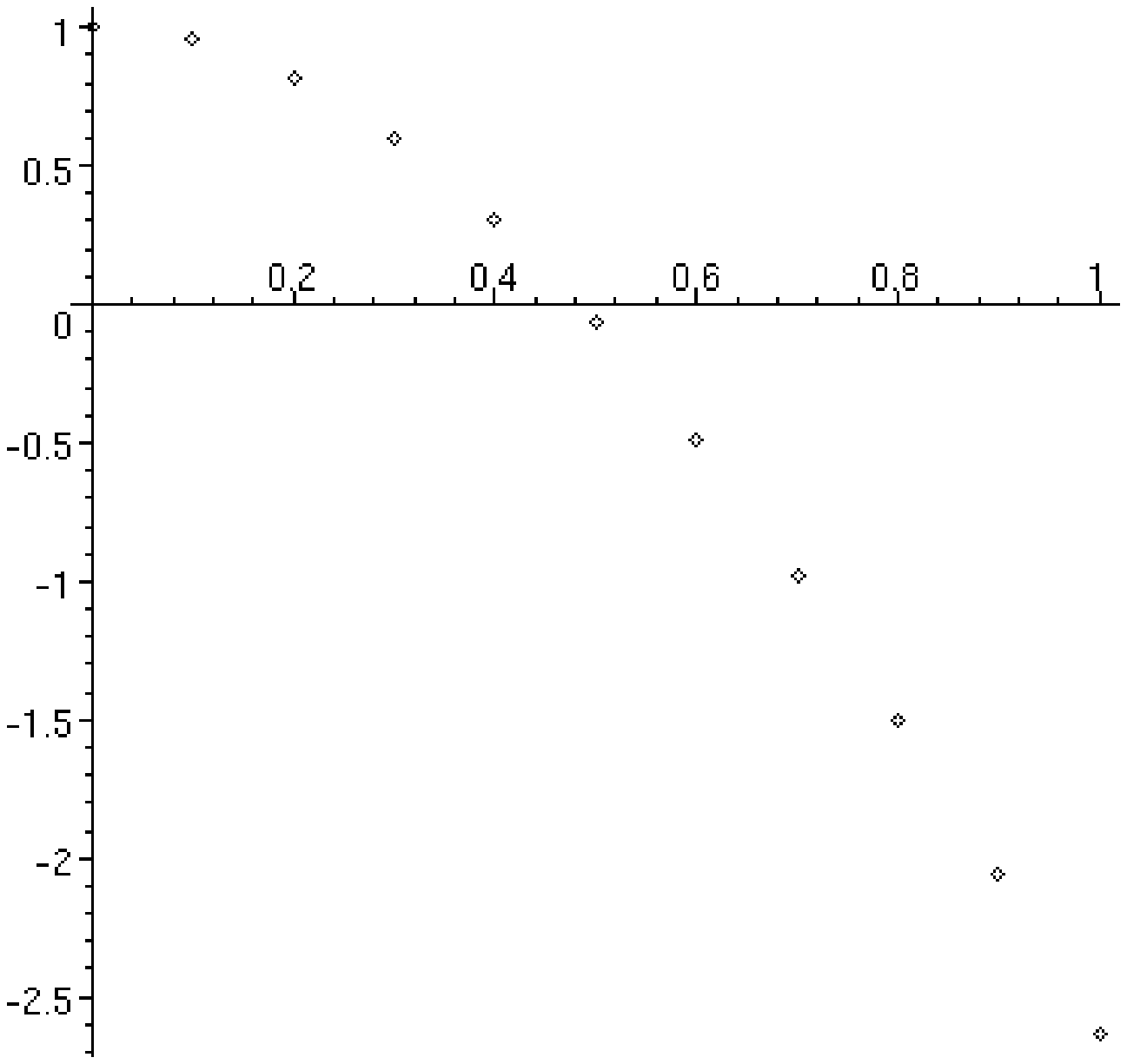}&\hspace{7pt}&\includegraphics[scale=0.28]{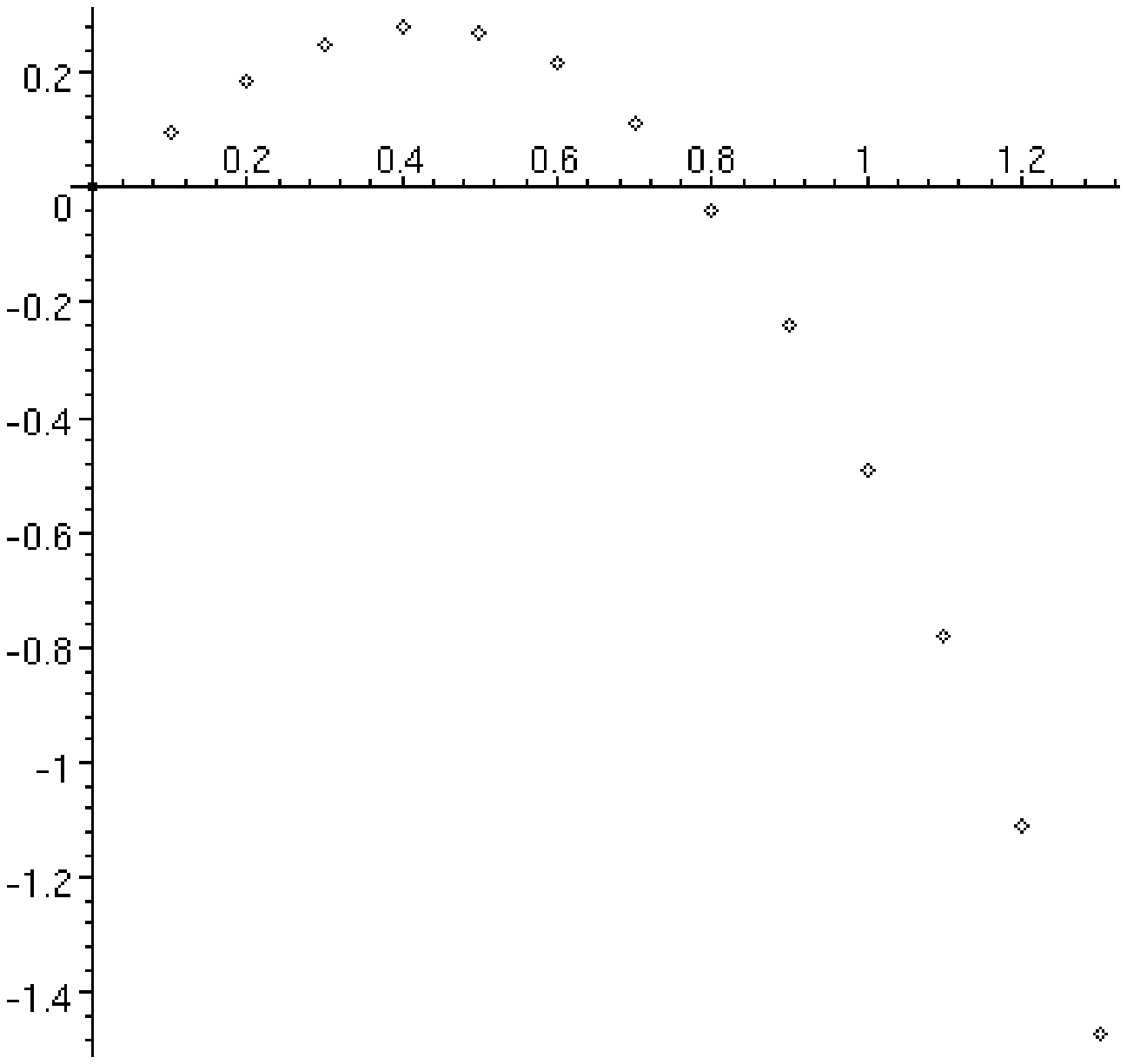} \\
\vspace{7pt}
\end{array}$
\end{center}
\caption{Prefactors corresponding to the first three excitations}
\label{Pq}
\end{figure}
Using the resummation technique already outlined, the prefactors corresponding to different levels of excitation may be plotted. The prefactors corresponding to the 1st, 2nd and 3rd excited states are shown in figure \ref{Pq} both on a large and small $x$ scale. Since these functions are either odd or even we restrict the domain to $x\ge0$. As would be expected the prefactor corresponding to the first excited state has one zero located at the origin. The 2nd excited state has two zeros and the 3rd excited state has 3 zeros when considered along the whole of the real axis. The tuning procedure did not provide a method for determining which level of excitation we have, just that we had determined an energy eigenstate. Using out resummation method to plot the prefactor we are able to confirm which energy eigenstate has been found by counting the number of nodes. The technique in section \ref{sec:hbar} provides an alternative method. We could use resummation in a $\hbar$ expansion to approximate $c_2$ or $c_3$ and then use the tuning method to determine the value more accurately. 
 
\begin{figure}
$\begin{array}{ccccc}
\includegraphics[scale=0.5]{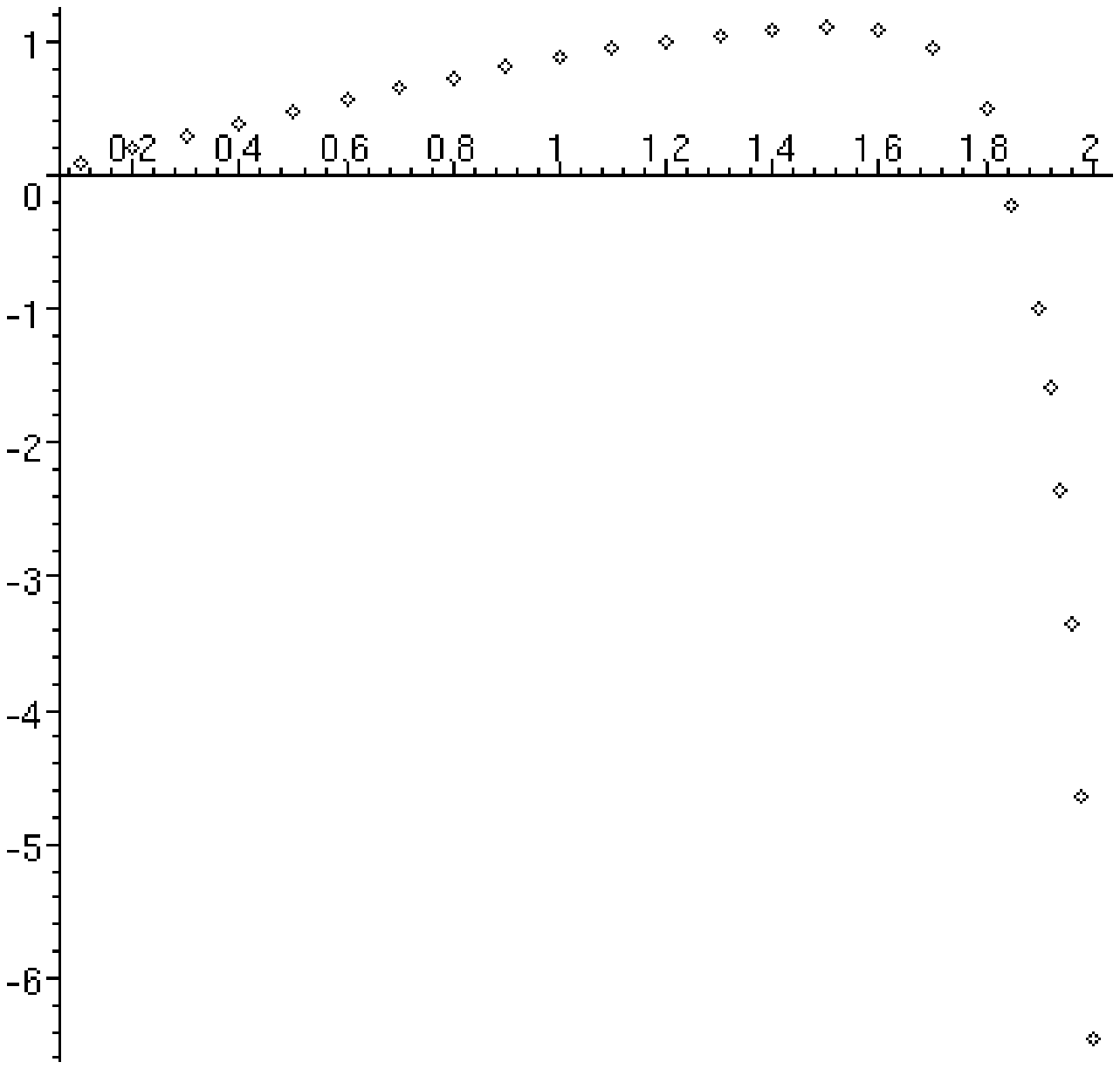}&\hspace{30pt}&\includegraphics[scale=0.5]{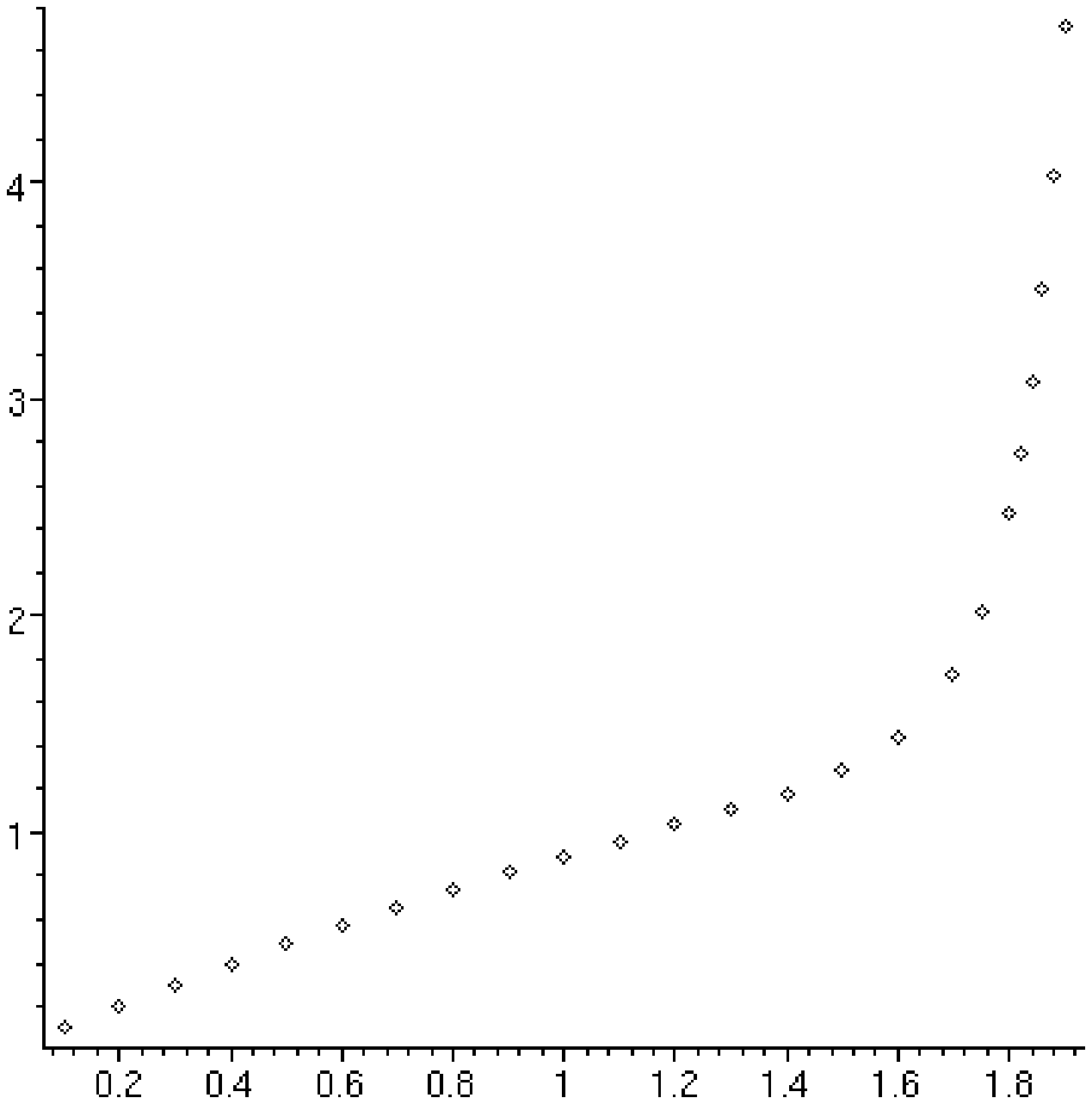}\\
$a) $c_1=-0.15$ $&&$b) $c_1=-0.14\\
\vspace{7pt}
\end{array}$
\caption{$P$ for values of $c_1$ either side of $e_1=-0.1458432840772$}
\label{P}
\end{figure}
\begin{figure}
$\begin{array}{ccccc}
\includegraphics[scale=0.5]{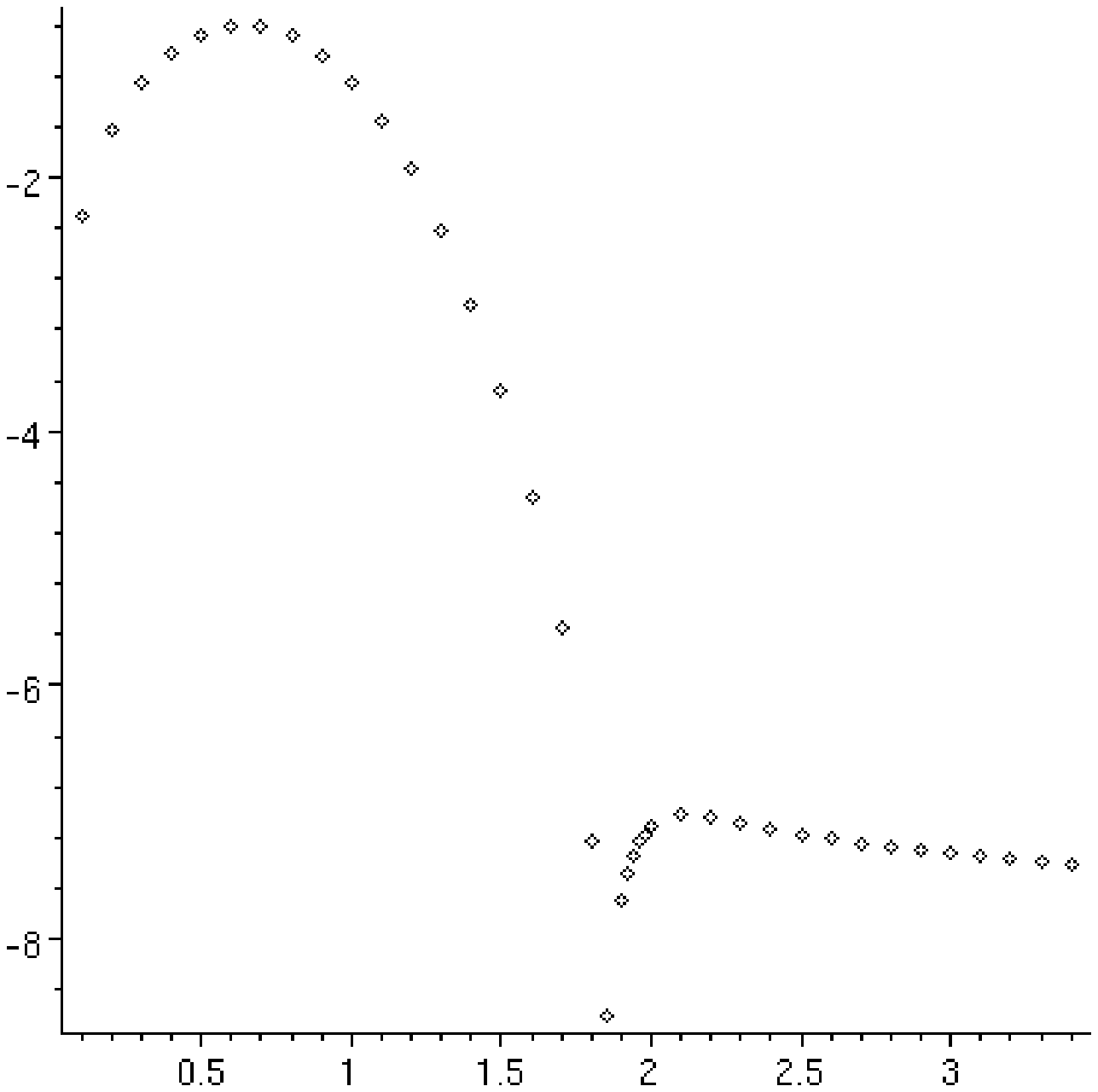}&\hspace{30pt}&\includegraphics[scale=0.5]{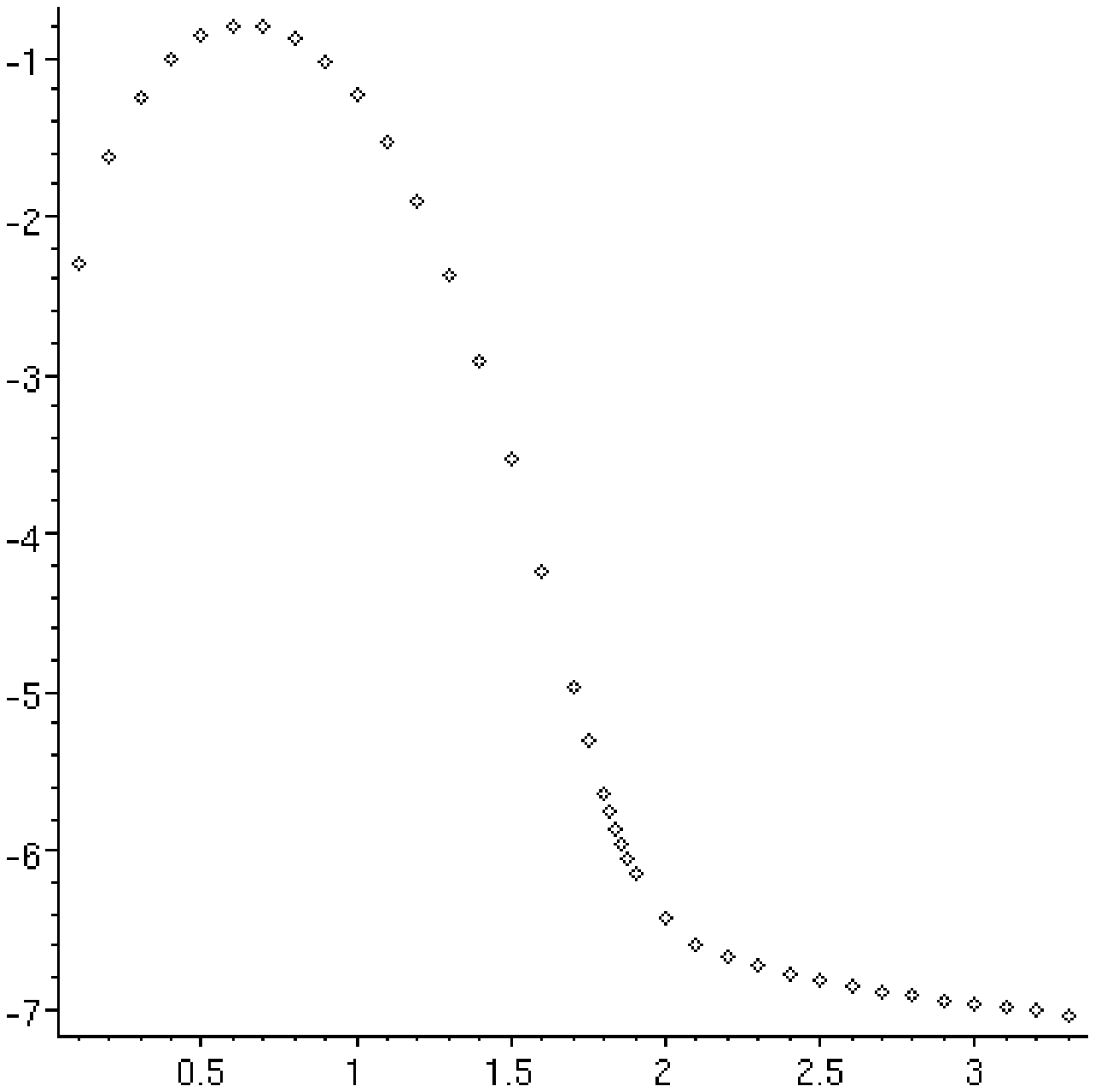}\\
$a) $c_1=-0.15$ $&&$b) $c_1=-0.14\\
\vspace{7pt}
\end{array}$
\vspace{-7pt}\caption{$Q$ for values of $c_1$ either side of $\tau_1\approx-0.1458432840772$}
\label{Q}
\end{figure}

We note that for large $x$ these prefactors asymptotically approach some constant value as predicted. For example the prefactor corresponding to some energies either side of the first excited state energy level are shown in figure \ref{P}. For $\tau_1<\tau<\tau_2$ we see that $P$ quickly becomes very large. This is similarly true for $\tau<\tau_1$ however this time $P$ becomes large in the negative direction. This appears not to be the result of a pole or cut otherwise for large $x$ the resummation technique would exhibit singularity contributions. Instead we find the prefactor is simply resumming to large values and these values are increasing in size rapidly. We test the hypothesis that we are observing prefactors with large $x$ behaviour determined by the second type of asymptotic behaviour in \eref{largeP} by plotting
\be
Q\equiv\log|P|-\frac{2}{3}\sqrt{g}x^3.
\ee
If the hypothesis is correct then we expect $Q$ to approach a constant for large $x$. We plot these graphs for those same values of $c_1$ as before and display them in figure \ref{Q}. As predicted these plots do flatten out for large $x$ at least on the scale of $\exp(x^3)$. This numerically supports our hypothesis.

\section{A Shifted Expansion}
\label{sec:shiftsc}
Let us momentarily remove $\alpha$ from $L(\lambda)$ by setting $\alpha=1$. We were unable to directly evaluate $L(\lambda)$ even having replaced $f(s)$ with the truncated asymptotic expansion. This was due to the $s-s_0$ denominator in the integrand which we expanded to give a sum of integrals of the form
\begin{equation}
\int_C ds\, \frac{e^{\lambda s}}{s^n}=\frac{\lambda^n}{\Gamma(n)}
\end{equation}
which are more easily evaluated. Unfortunately the expansion of this denominator required summing over two variables in $L_N(\lambda)$. We truncated the expansion of the $(s-s_0)^{-1}$ term to a relatively high order to avoid complications arising from a truncated form of this expansion. This effectively meant we had to compute a series involving a sum of $NP$ terms. With our chosen $N$ and $P$ this amounted to $1500$ terms. 

Instead we shall briefly investigate the possibility of shifting $s\to s+s_0$ in $L(\lambda)$ and then using the truncated expansion of $f(s)$ to approximate $L(\lambda)$. That is we write
\begin{equation}
f(s+s_0)=\sum_{n=0}^\infty\frac{c_n}{(s+s_0)^n}=\sum_{n=0}^\infty \frac{\tilde c_n}{s^n}
\end{equation}
by expanding the $(s+s_0)^{-n}$ terms in powers of $s_0/s$. Our new coefficients $\tilde c_n$ would then be dependent on $s_0$ so a separate expansion would be required for each $s_0$ we try to evaluate.  

We then truncate the new expansion and substitute into $L(\lambda)$. Having done this we reinsert the parameter $\alpha$ with the substitution $s\to s^\alpha$.
\begin{equation}
\label{LNnew}
L_N(\lambda)=\sum_{n=0}^N \tilde c_n\frac{\lambda^{\alpha n}}{\Gamma(\alpha n)}.
\end{equation}
The advantage of this type of expansion is that we no longer need to worry about a double summation truncated to order $N$ and $P$. This reduces the computational time to evaluate the series but also prevents the problems encountered in \ref{sec:hbar} as a result of truncation in $P$. 

We should note however that the $\alpha$ parameter in the original $L_N$ \eref{LN} is different to the $\alpha$ in \eref{LNnew}. In the original formulation $\alpha$ caused singularities to be rotated about the origin. This is still the case in the new formulation however we now have a different origin since $s$ has been shifted. It will depend on the location of singularities as to which method will allow a larger $\alpha$ and therefore better dampening of pole contributions.  

\begin{figure}
\begin{center}
\begin{tabular}{|c|c|c|c|c|c|c|}
\hline
$\mathbf{\hbar}$&$\mathbf{\alpha_M}$&$\mathbf{\lambda_M}$&$\mathbf{b_2}$&$\mathbf{g_{\textit{est}}}$&$\mathbf{g_\textit{exact}}$&\bf{Err ($\mathbf{\%}$)}\\
\hline
0.05&1.0000&0.5166&1.0289943&0.0578329&0.0578320&0.002\\
0.10&1.0040&0.5631&1.0545995&0.1345734&0.1345815&0.006\\
0.15&1.0627&0.7669&1.0791362&0.2372547&0.2370183&0.100\\
0.20&1.0971&0.9049&1.1021792&0.3763874&0.3755570&0.221\\
0.25&1.1303&1.0487&1.1244110&0.5689119&0.5667201&0.387\\
0.30&1.1569&1.1569&1.1457268&0.8426714&0.8379430&0.564\\
0.35&1.1832&1.3033&1.1665395&1.2471800&1.2375945&0.775\\
0.40&1.2060&1.4223&1.1866922&1.8763800&1.8579267&0.993\\
0.45&1.2264&1.5340&1.2062905&2.9298231&2.8946902&1.214\\
0.50&1.2435&1.6317&1.2252384&4.9010160&4.8319442&1.429\\
0.55&1.2625&1.7442&1.2441147&9.3469363&9.1926555&1.678\\
0.60&1.2777&1.8371&1.2622842&23.9503681&23.5025564&1.905\\
0.65&1.2941&1.9401&1.2803813&210.5377762&206.0985278&2.154\\
\hline
\end{tabular}
\end{center}
\caption{Results of resummation in the shifted semi classical expansion}
\label{tab:hshifted}
\end{figure}
We complete our resummation process using the shifted $\hbar$ expansion and present the results in figure \ref{tab:hshifted}. We note that the errors are considerably larger than in the previous method. So whilst this method does have some advantages it is clearly less efficient for the semi classical expansion. We attribute this to the different geometry involved when rotating poles by varying $\alpha$. It is clear that in the shifted method we have smaller values of $\alpha_M$ and therefore $\lambda_M$ which results in less dampening of any pole contributions. Although this method is inferior for the expansion of $b_2$ we should note that depending on the location of singularities it may prove to be better in other expansions and therefore should not be ignored.

We could question however whether it is fair to compare this method with $N=30$ to the previous method which effectively had $NP=1500$ terms. In the shifted example we could afford to take more terms given the reduced computational power required to perform the resummation. We note however that in some perturbative expansions calculating an expansion to higher orders can often be the limiting factor.

\section{Error Estimates}
There are two main sources of error in our prescription for evaluating $f(s_0)$. The first is due to $L_N(\lambda)$ only being an approximation to $L(\lambda)$. We require $L_N(\lambda)$ to differ from $L_{N-1}(\lambda)$ by no more than say $\sigma$ percent. Therefore we can think of $\sigma$ as being an error. Smaller $\sigma$ requires evaluating $L(\lambda)$ for a smaller $\lambda$ which has the unwanted side effect of reducing the exponential dampending of any singularities of $f(s)$ in the left half plane. This is the second source of error. We shall outline a method for estimating the error in $L(\lambda)$ as a result of singularity contributions.

Consider a pole of $f(s)$ located at $s_p$. We will only be working with asymptotic expansions for which all coefficients, $a_n$ are real. This implies that for a pole to exist at $s_p$ we must necessarily have a pole at the conjugate location $s_p^*$. We shall for now assume that $f(s)$ has just one pair of poles in the left half $s$ plane but is analytic on the remainder of the complex plane. The pair of conjugate poles gives a contribution to \eref{L(lambda)} of the form
\begin{equation}
\label{cor}
c\cos(\lambda y_p+\nu)e^{\lambda(x_p-s_0^{1/\alpha})}
\end{equation}
where $s_p^{1/\alpha}$ is split into real and imaginary parts $s_p^{1/\alpha}=x_p+iy_p	$ and $c$, $\nu$ are real numbers. In the large $\lambda$ limit this contribution approaches zero provided $\alpha$ is not taken so large as to rotate the poles into the right half plane. For a general $\alpha$ the correction \eref{cor} looks like an oscillating curve either growing or decreasing in amplitude. For $x_p>s_0^{1/\alpha}$ the oscillations are growing and for $x_p<s_0^{1/\alpha}$ the oscillations are being damped. We will use this to tune $\alpha$ until the oscillations are fixed in amplitude, say $\alpha=\alpha_T$ at which point $x_p=s_0^{1/\alpha}$. Having determined $x_p$ we determine $y_p$ by calculating the period of oscillations. This is best achieved by considering the $\lambda$ derivative of $L_N(\lambda)$ and looking for zeros with $\alpha=\alpha_T$. The phase $\nu$ and amplitude $c$ are then easily calculated by fixing the location of zeros in the derivative of $L_N(\lambda)$ and also ensuring the correct amplitude of the sinusoidal curve. Now with $\alpha=\alpha_T$ the pole contribution is
\begin{equation}
\frac{e^{\lambda y_p i}}{iy_p}\rho_T+\textit{cc}=c\cos(\lambda y_p+\nu)
\end{equation}
where $\rho_T$ is the residue of $f(s^{\alpha_T})$ at $s_p^{1/\alpha_T}$ and $\textit{cc}$ denotes the complex conjugate. So we have determined $\rho_T$ numerically in terms of the parameters $c$, $y_p$ and $\nu$.

We want to evaluate $L(\lambda)$ (or $L_N(\lambda)$) when $\alpha=\alpha_M$, the maximal value of $\alpha$ for which the curve is still monotonic. With this value of $\alpha$ the pole contribution is
\begin{equation}
\label{Lerror}
\rho_M\frac{e^{\lambda\left(s_c-s_0^{1/\alpha_M}\right)}}{s_c-s_0^{1/\alpha_M}}+\textit{cc}
\end{equation}
where
\begin{equation}
s_c\equiv\left(s_0^{1/\alpha_T}+iy_p\right)^{\alpha_T/\alpha_M}
\end{equation}
and $\rho_M$ is the residue of $f(s^{\alpha_M})$ at $s_c$. We can relate $\rho_T$ and $\rho_M$ via
\begin{equation}
\rho_T=\rho_M\frac{\alpha_M}{\alpha_T}\frac{s_c}{s_p^{1/\alpha_T}}
\end{equation}
and therefore can calculate the pole contribution from \eref{Lerror}. We think of $\sigma$ as being the error in approximating $L(\lambda)$ with $L_N(\lambda)$ which in this case was chosen to be $\sigma=10^{-3}\%$ and \eref{Lerror} as being the correction to $L(\lambda)$ as an approcimation to $f(s_0)$ as a result of a pole pair contribution.

In general $f(s)$ will not have just one pair of poles but a number of poles and cuts. The contributions from these singularities will be exponentially weighted with singularities lying furthest to the right being most dominant. Due to this exponential weighting a cut contribution will look like a pole dominated by the right most end of the cut. We can therefore apply our procedure assuming the existence of just one pole pair and then interprate the correction \eref{Lerror} as being the dominant or leading order correction. We note however that the right most end of a cut can actually change as it is rotated by increasing $\alpha$. The left and right ends of a cut may actually switch if $\alpha$ becomes too large. In this case the error estimate may not be the most dominant but will be a more minor correction.

In order to apply our technique we will require sufficient terms to ensure at least two peaks or troughs for $\alpha=\alpha_T$. These must exist within a region where $L_N(\lambda)$ is a good approximation to $L(\lambda)$. Clearly $N=30$ in the Bender Wu expansion is insufficient for large couplings as $L_N(\lambda)$ did not approximate $L(\lambda)$ sufficiently well. We also note that with the semi classical expansion, the errors where of the order $10^{-4}\%$. So in this case the dominant error is provided by $\sigma$. In fact the exact values lie between $L_{30}$ and $L_{29}$ in this expansion so $\sigma$ is actually an error bound. It would however be pointless to calculate the error correction due to singularities without first reducing $\sigma$ (or $N$). We will instead apply this technique to the shifted semi classical expansion (section \ref{sec:shiftsc}). In this technique we found the actual error to be much greater than $\sigma$. 

\begin{figure}
\begin{center}
\begin{tabular}{|c|c|c|c|c|c|c|}
\hline
$\mathbf{g_\textit{best}}$&$\mathbf{\alpha_T}$&$\mathbf{g_\textit{corr}}$&\bf{Resum Err}&\bf{Corrected Err}&\bf{Relative Err}\\
\hline
0.057832&1.0000&0.057833&0.00165&0.00002&90\\
0.134581&1.0040&0.134613&0.00600&0.00023&26\\
0.237018&1.0627&0.237378&0.09976&0.00152&66\\
0.375557&1.0971&0.376543&0.22112&0.00263&84\\
0.566720&1.1303&0.568924&0.38674&0.00389&99\\
0.837943&1.1569&0.842246&0.56430&0.00514&110\\
1.237595&1.1832&1.243168&0.77452&0.00450&172\\
1.857927&1.2060&1.866679&0.99322&0.00471&211\\
2.894690&1.2264&2.908286&1.21370&0.00470&258\\
4.831944&1.2435&4.864863&1.42948&0.00681&210\\
9.192656&1.2625&9.237419&1.67830&0.00487&345\\
23.502556&1.2777&23.666373&1.90537&0.00697&273\\
206.098528&1.2941&207.519940&2.15394&0.00690&312\\
\hline
\end{tabular}
\end{center}
\caption{Corrected couplings due to dominant singularity contributions}
\label{tab:error}
\end{figure}

We subtract the dominant singularity correction and list the results ($g_{\textit{corr}}$) in figure \ref{tab:error}. The percentage errors after resummation and percentage errors having subtracted the dominant singularity contribution are listed. The relative error is the ratio of these two errors and indicates that the percentage error has improved by a factor of between $90$ and $312$. We expect the error as a result of singularities is greater for larger $\hbar$ and this is reflected in these increasing corrections for higher $\hbar$.

\section{Summary}
The most impressive application of the modified Borel resummation technique in this paper is not to look for an approximate value of a resummed expansion but instead to use the analytic properties of a function to tune for the correct boundary condition (section \ref{tuning}). This method was first outlined in \cite{Leonard:2007dj} and in this paper we have given numerical evidence to explain why the technique works. Due to the arbitrary levels of accuracy that may be obtained through this method it is worthy of futher study. It has already been applied to anharmonic oscillators with higher order potentials however we believe it may be adaptable to solve other types of problems.  

The concept of direct resummation is also important. We have used it to numerically investigate the location of zeros in solutions to the anharmonic differential equation and therefore explain how the tuning technique selects the solution with the correct boundary condition. We found that when the correct boundary condition is not satisfied the logarithm of the wavefunction contains singularities either on the real axis or in the complex plane depending on whether the energy chosen is too large or too small. The existence of these singularities causes an easily observable contribution though the modified Borel resummation procedure in the form of a rapidly increasing or decreasing curve. 

In other physical problems we are not always in a position to apply a tuning type method where we look for the correct large $x$ behaviour. We have shown that the modified Borel summation technique can produce reasonably accurate results from perturbative expansions that are ordinarily divergent. This has been shown by considering two different perturbative expansions of the wavefunction. The two expansions have different singularity locations and residues but in both cases these are the half $x$ plane with negative real part. The semi classical expansion in $\hbar$ was particularly useful because it separates the $g^{1/3}$ type behaviour for large $g$ from the expansion we intend to resum. Therefore we get much better results for this part of the spectrum than if we used the Bender Wu coupling expansion.

We have taken the modified Borel technique and improved it by comparing two different methods for approximating the contour integral. Both techniques have their advantages depending on the location of any singularity contributions and how changing the $\alpha$ parameter causes them to move through the complex plane. We have investigated the major sources of error and improved the accuracy of the procedure by subtracting the leading order corrections. Also, in contrast to standard Borel summation we do not require an analytic continuation of the Borel sum since this is encoded in the contour integral.

\end{document}